\documentclass[11pt,a4paper]{article}
\usepackage[utf8]{inputenc}
\usepackage[T1]{fontenc}
\usepackage{lmodern}
\usepackage{amsmath,amssymb,amsthm,bm}
\usepackage{graphicx}
\usepackage{microtype}
\usepackage[margin=2.6cm]{geometry}
\usepackage{xcolor}
\usepackage{enumitem}
\setlist[itemize]{topsep=3pt,itemsep=1pt,parsep=0pt}
\usepackage[colorlinks=true,linkcolor=blue!50!black,citecolor=blue!50!black,urlcolor=blue!50!black]{hyperref}

\newtheorem{theorem}{Theorem}
\newtheorem{lemma}{Lemma}
\newtheorem{proposition}{Proposition}

\theoremstyle{definition}
\newtheorem{definition}{Definition}

\newcommand{\Pp}{\mathcal{P}}
\newcommand{\E}{\mathbb{E}}
\newcommand{\R}{\mathbb{R}}
\newcommand{\C}{\mathbb{C}}
\newcommand{\N}{\mathbb{N}}
\newcommand{\Exp}{\mathrm{Exp}}
\newcommand{\Tr}{\mathrm{Tr}\,}
\newcommand{\vol}{\mathrm{vol}}
\newcommand{\rpar}{\mathbf{r}^{\parallel}}
\newcommand{\Rinf}{\R_+^{\infty}}

\title{The infinite-dimensional extended Bloch representation}
\author{Massimiliano Sassoli de Bianchi\\[2pt]
{\small Center Leo Apostel for Interdisciplinary Studies, Brussels Free University}\\
{\small Krijgskundestraat 33, 1160 Brussels, Belgium}\\
{\small E-Mail: \texttt{msassoli@vub.ac.be}}}
\date{\small November 24, 2026}

\begin{document}
\maketitle

\begin{abstract}
\noindent In the extended Bloch representation (EBR) of quantum mechanics, the Born probabilities of a measurement with $N$ outcomes are the relative volumes of $N$ regions of a simplex-shaped abstract membrane, whose points represent the ``hidden'' measurement interactions. However, when $N\to\infty$, the volumes vanish and the uniform membrane concentrates at the center of the unit ball. We show that this problem can be solved by renormalization: if the radius of the generalized Bloch ball grows like $N$, the distances of the measurement interactions from the facets of the simplex become independent exponential random variables. This is the only growth with a non-degenerate limit, and the limit membrane, which cannot live in the limit of the generalized Bloch space without privileging some of the outcomes, is the projective limit of the finite-dimensional membranes and, once renormalized, the only distribution compatible with the Born rule. For a pure point spectrum, the regions of the membrane become cones, the particle representing the state recedes to infinity, and the lines along which the membrane collapses become parallel. For a continuous spectrum, a further magnification yields a Poisson point process and the Born rule for probability densities. We also show that the uniform membrane is the only one compatible with the Born rule in finite dimension, and that the mechanism can be read, in every dimension, as a race between independent exponential clocks.
\end{abstract}
{\raggedright\noindent{\bf Keywords}:
Extended Bloch representation; Hidden measurements; Born rule; Uniform distribution on a simplex; Poisson point processes; Exponential race\par}

\section{Introduction}

\begin{figure}[tp]
\centering
\includegraphics[width=0.8\textwidth]{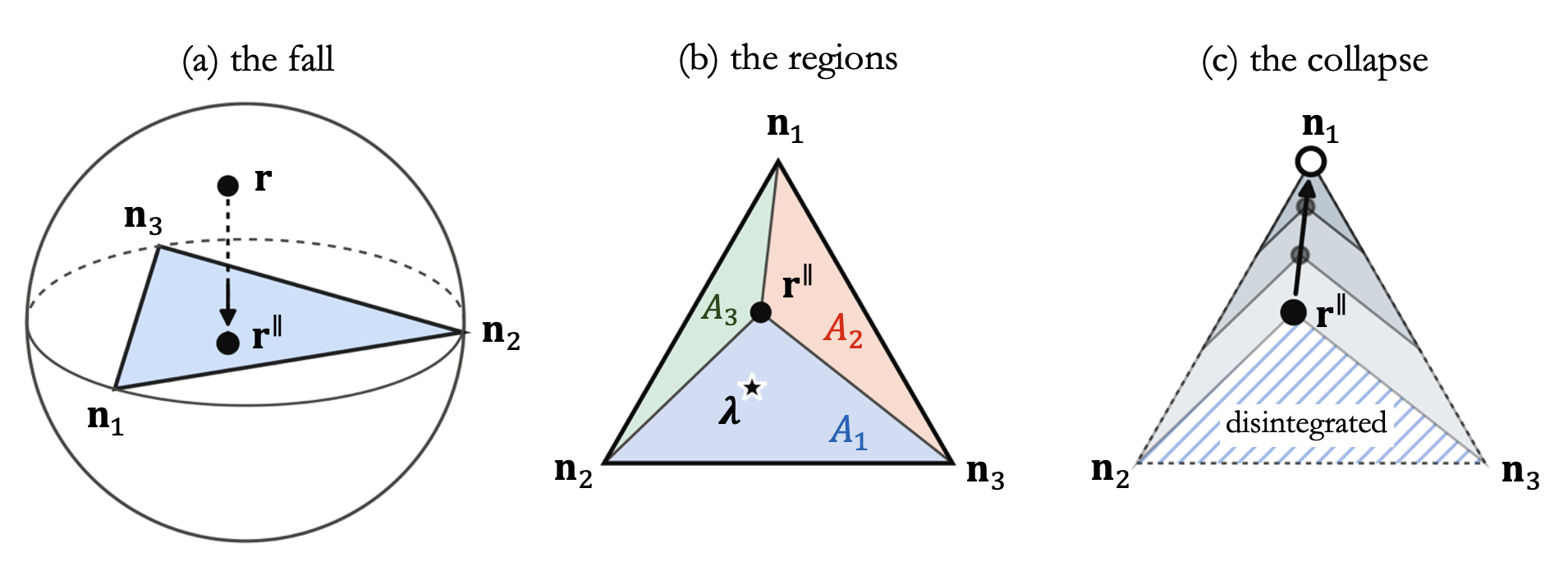} 
\caption{The measurement process of the EBR for $N=3$. (a) The point particle representing the state, initially at $\mathbf{r}$, falls orthogonally onto the measurement simplex, a triangle with vertices $\mathbf{n}_1$, $\mathbf{n}_2$, $\mathbf{n}_3$, reaching the point $\rpar$ whose barycentric coordinates are the Born probabilities, here $p_1=0.5$, $p_2=0.3$ and $p_3=0.2$. The drawing is schematic: for $N=3$ the generalized Bloch ball is eight-dimensional, and only the plane of the triangle and one direction orthogonal to it are represented. (b) The particle divides the triangle into three regions: $A_i$ is the triangle having the particle as its apex and the edge opposite to the vertex $\mathbf{n}_i$ as its base, and the areas of the three regions are in the proportions $p_1:p_2:p_3$. The membrane starts disintegrating at a point $\bm{\lambda}$, here located in $A_1$. (c) The whole region $A_1$ disintegrates (hatched), the membrane loses its anchorage to the vertices $\mathbf{n}_2$ and $\mathbf{n}_3$ and, by contracting (shaded shapes), draws the particle to the vertex $\mathbf{n}_1$, which represents the eigenstate $P_{a_1}$: the outcome of the measurement is $a_1$.}
\label{fig:process}
\end{figure}

The extended Bloch representation (EBR) of quantum mechanics~\cite{AS2014,AS2016a,AS2017a} generalizes the three-dimensional Bloch sphere~\cite{Aerts1986} to quantum entities with a Hilbert space of arbitrary finite dimension $N$. The states are represented by the points of a convex region of a unit ball of $\R^{N^2-1}$, the \emph{generalized Bloch ball}, and a measurement with $N$ outcomes by a regular $(N-1)$-dimensional simplex inscribed in it, the \emph{measurement simplex}, which can be pictured as an abstract elastic and disintegrable membrane. The measurement is then a process in which the point particle representing the state first ``falls'' orthogonally onto the membrane, and is then drawn to one of its vertices when the membrane starts disintegrating at some unpredictable point $\bm{\lambda}$ and collapses. If the points $\bm{\lambda}$, which represent the possible ``hidden'' measurement interactions between the entity and the apparatus, are uniformly distributed, the probability of each outcome is the relative Lebesgue measure of the region $A_i$ of the simplex associated with it, which coincides with the Born probability. The process is illustrated in Fig.~\ref{fig:process}, for $N=3$.

It is then natural to ask what becomes of this description when $N\to\infty$, since quantum entities as simple as an electron require an infinite-dimensional Hilbert space. In~\cite{AS2014} (Sec.~14), the transition probabilities of an infinite-dimensional entity were shown to be limits of finite-dimensional ones, so that a hidden-measurement interpretation can be advocated for them, but without constructing the asymptotic representation. The obstacle to a direct construction was identified in~\cite{AS2017b}: the volumes of a ball of fixed radius and of an inscribed simplex tend to zero extremely fast as the dimension increases. Since actual measurements have a finite number of outcomes, possibly infinitely degenerate, for which the EBR applies without modification, the question of whether an indeterministic process ``describable as a weighted symmetry breaking over the available hidden-measurement interactions''~\cite{AS2017b} takes place on an infinite-dimensional simplex was answered only for a finite number of outcomes.

The purpose of this article is to answer this question in the affirmative also for an infinite number of outcomes. For this, we consider the standard treatment of the continuum limit of the eigenvectors of position: if position is approximated by an observable whose eigenvalues form a lattice of spacing $\epsilon$, the components $\langle v_x|\psi\rangle$ of a state along the normalized eigenvectors $|v_x\rangle$ tend to zero with $\epsilon$, and one obtains the wave function only if the eigenvectors are renormalized, $|x\rangle=|v_x\rangle/\sqrt{\epsilon}$, at the price of their norm becoming infinite. The vanishing volumes of the EBR are a problem of the same kind, and they have the same solution: the radius of the generalized Bloch ball should grow with the dimension.

After recalling the EBR (Section~\ref{sec:EBR}), we show that its three length scales, the radius of the ball, that of the sphere inscribed in the simplex and the typical distance of the measurement interactions from the center, are in the proportions $1:\frac{1}{N-1}:\frac{1}{\sqrt{N+1}}$, so that the unit ball keeps the states but loses the measurement interactions (Section~\ref{sec:unit}). With the radius $R_N=N-1$ (Section~\ref{sec:renormalized}), the distances of the measurement interaction from the facets become independent exponential variables,  
and, as long as all the outcomes are treated on the same footing, this limit necessarily lies outside the limit of the generalized Bloch space (Section~\ref{sec:limit}). The limit object, the \emph{infinite membrane}, is the projective limit of the uniform membranes (Section~\ref{sec:infinite-membrane}). For a pure point spectrum, its regions are cones whose measures are the Born probabilities, and it is the only distribution with this property (Section~\ref{sec:purepoint}). For a continuous spectrum, a second magnification yields a Poisson point process (Section~\ref{sec:continuous}). 
Finally, we discuss the two normalizations that separate the states from the measurement interactions, and observe that the uniform membrane can be read, in every dimension, as a race between exponential clocks (Section~\ref{sec:discussion}).

\section{The EBR in a nutshell}
\label{sec:EBR}

In this section we recall the elements of the EBR that will be needed in the following, together with two elementary lemmas on the regions of the membrane and on the uniform distribution on a simplex.

\subsection{States and measurement interactions}

Let $\mathcal{H}=\C^N$, and let $\Lambda_1,\dots,\Lambda_{N^2-1}$ be a choice of generators of $SU(N)$, i.e., of traceless self-adjoint matrices such that $\Tr\Lambda_i\Lambda_j=2\delta_{ij}$. Every operator-state (density operator) can be written as
\begin{equation}
D(\mathbf{r})=\frac{1}{N}\left(\mathbb{I}+c_N\,\mathbf{r}\cdot\bm{\Lambda}\right),\quad c_N=\sqrt{\frac{N(N-1)}{2}}
\label{eq:state}
\end{equation}
where $\mathbf{r}$ is a vector of the unit ball $B_1(\R^{N^2-1})$. The vector-states (rank-one projection operators) correspond to unit vectors, but for $N>2$ only a convex portion of the ball is filled with states~\cite{AS2014,KimuraKossakowski2005}. 
We have: 
\begin{equation}
\Tr D(\mathbf{r})D(\mathbf{s})=\frac{1}{N}\left[1+(N-1)\,\mathbf{r}\cdot\mathbf{s}\right]
\label{eq:transition}
\end{equation}
Indeed, $\Tr(\mathbf{r}\cdot\bm{\Lambda})(\mathbf{s}\cdot\bm{\Lambda})=\sum_{i,j}r_is_j\Tr\Lambda_i\Lambda_j=2\,\mathbf{r}\cdot\mathbf{s}$, so that,
the $\Lambda_i$ being traceless, $\Tr D(\mathbf{r})D(\mathbf{s})=\frac{1}{N^2}\big(N+2c_N^2\,\mathbf{r}\cdot\mathbf{s}\big)$, and $2c_N^2=N(N-1)$.

Let $A=\sum_{i=1}^Na_iP_{a_i}$ be an observable with a non-degenerate spectrum, and let $\mathbf{n}_i$ be the unit vectors representing its eigenstates, $P_{a_i}=D(\mathbf{n}_i)$. By~(\ref{eq:transition}), $\mathbf{n}_i\cdot\mathbf{n}_j=-\frac{1}{N-1}$, for $i\neq j$: the $\mathbf{n}_i$ are the vertices of a regular $(N-1)$-dimensional simplex $\triangle_{N-1}$, inscribed in the unit ball and centered at the origin, since $\sum_i\mathbf{n}_i=\mathbf{0}$, by~(\ref{eq:state}) and $\sum_iP_{a_i}=\mathbb{I}$. Every point of the simplex can be written, in a unique way, as
\begin{equation}
\bm{\lambda}=\sum_{j=1}^N\lambda_j\mathbf{n}_j,\quad\lambda_j\geq 0,\quad\sum_{j=1}^N\lambda_j=1
\label{eq:barycentric}
\end{equation}
and the $\lambda_j$ are its \emph{barycentric coordinates}. 

\begin{figure}[tp]
\centering
\includegraphics[width=0.8\textwidth]{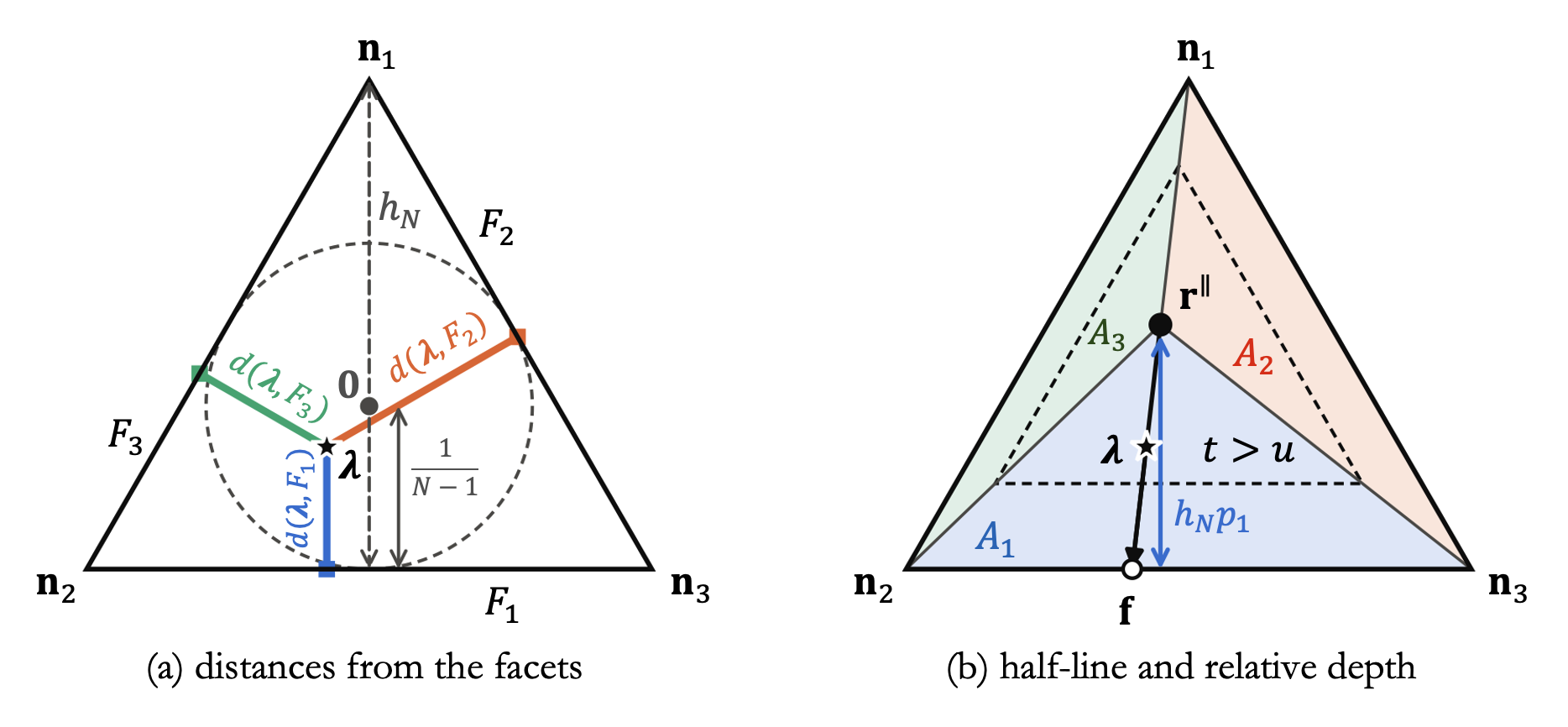} 
\caption{The geometry of the measurement simplex for $N=3$. (a) 
The distances $d(\bm{\lambda},F_i)=h_N\lambda_i$ of the disintegration point from the facets are proportional to its barycentric coordinates. The height of the simplex is $h_N=\frac{N}{N-1}=\frac{3}{2}$, and the inscribed circle (dashed), centered at the origin, has radius $\frac{1}{N-1}=\frac{1}{2}$. (b) The half-line issued from the particle and passing through $\bm{\lambda}$ leaves the simplex at the point $\mathbf{f}$ of $F_1$, which shows that $\bm{\lambda}\in A_1$ [Lemma~\ref{lem:regions}(b)]. The relative depth of $\bm{\lambda}$ is the ratio between its distance from $\mathbf{f}$ and that of the particle, here $t(\bm{\lambda})=\frac{1}{2}$. The dashed triangle is the image of the simplex under the homothety of center $\rpar$ and ratio $1-u$, and it contains the points with relative depth larger than $u$; see Lemma~\ref{lem:regions}(c).}
\label{fig:distances}
\end{figure}

Let $F_i$ be the \emph{facet} of the simplex opposite to the vertex $\mathbf{n}_i$, i.e., the set of the points with $\lambda_i=0$. Since $(\mathbf{n}_j-\mathbf{n}_k)\cdot\mathbf{n}_i=0$, for $j,k\neq i$, the facet $F_i$ is orthogonal to $\mathbf{n}_i$, and since its centroid is $\frac{1}{N-1}\sum_{j\neq i}\mathbf{n}_j=-\frac{1}{N-1}\mathbf{n}_i$, it lies in the hyperplane $\mathbf{x}\cdot\mathbf{n}_i=-\frac{1}{N-1}$. Hence, the distance $d(\bm{\lambda},F_i)$, of $\bm{\lambda}$ from $F_i$, is $\bm{\lambda}\cdot\mathbf{n}_i+\frac{1}{N-1}$, and since $\bm{\lambda}\cdot\mathbf{n}_i=\lambda_i-\frac{1-\lambda_i}{N-1}=\frac{N\lambda_i-1}{N-1}$, we have:
\begin{equation}
d(\bm{\lambda},F_i)=h_N\,\lambda_i
\label{eq:distance}
\end{equation}
where $h_N=1+\frac{1}{N-1}=\frac{N}{N-1}$ is the height of the simplex. In particular, the center of the simplex ($\lambda_j=\frac{1}{N}$, for all $j$) is at distance $\frac{1}{N-1}$ from all the facets: this is the radius of the sphere inscribed in the simplex, which is also the largest sphere centered at the origin that is completely filled with states~\cite{AS2014,KimuraKossakowski2005}; see Fig.~\ref{fig:distances}(a).

According to the EBR, the measurement of $A$ on an entity in the state $D(\mathbf{r})$ is a process made of the following stages~\cite{AS2014}. First, the point particle representing the state moves orthogonally to the simplex, from $\mathbf{r}$ to the point
\begin{equation}
\rpar=\sum_{i=1}^Np_i\,\mathbf{n}_i,\quad p_i=\Tr D(\mathbf{r})P_{a_i}=\frac{1}{N}\left[1+(N-1)\,\mathbf{r}\cdot\mathbf{n}_i\right]
\label{eq:Born}
\end{equation}
whose barycentric coordinates are the Born probabilities: since $\mathbf{r}-\rpar$ is orthogonal to the simplex, $\mathbf{r}\cdot\mathbf{n}_i=\rpar\cdot\mathbf{n}_i=\frac{Np_i-1}{N-1}$, as above, and solving for $p_i$ gives~(\ref{eq:Born}), which by~(\ref{eq:transition}) is $\Tr D(\mathbf{r})P_{a_i}$. This deterministic stage describes a decoherence-like process [Fig.~\ref{fig:process}(a)]. The simplex is then viewed as an abstract membrane, made of a disintegrable elastic substance, to which the particle remains attached, and which it divides into $N$ regions $A_i$, which are the convex hulls of $\rpar$ and  the facets $F_i$; see Fig.~\ref{fig:process}(b). At some moment, the membrane breaks at an unpredictable point $\bm{\lambda}$. If $\bm{\lambda}\in A_i$, the whole region $A_i$ disintegrates, the membrane loses its anchorage to all the vertices but $\mathbf{n}_i$, and by contracting it draws the particle to $\mathbf{n}_i$: the outcome is $a_i$, and the final state $P_{a_i}$; see Fig.~\ref{fig:process}(c).

The disintegration points $\bm{\lambda}$ describe the different ``hidden'' measurement interactions that can be actualized between the entity and the measuring apparatus. If they are uniformly distributed in the simplex, the probability of the outcome $a_i$ is
\begin{equation}
\mu_N(A_i)=\frac{\vol[A_i]}{\vol[\triangle_{N-1}]}=p_i
\label{eq:volumes}
\end{equation}
where $\vol$ denotes the $(N-1)$-dimensional Lebesgue measure, and $\mu_N$ the uniform probability measure on $\triangle_{N-1}$. The equality~(\ref{eq:volumes}), which is the Born rule, follows from the fact that $A_i$ and the simplex are two pyramids with the same base $F_i$, and apexes $\rpar$ and $\mathbf{n}_i$, whose distances from $F_i$ are, by~(\ref{eq:distance}), $h_Np_i$ and $h_N$ [Fig.~\ref{fig:distances}(b)]. Note that the regions $A_i$ fill the simplex and only overlap on sets of zero volume, in accordance with $\sum_ip_i=1$.

\subsection{Regions and uniformly distributed points}

The following two elementary lemmas will be used throughout the article. The first one characterizes the regions $A_i$ in terms of barycentric coordinates. The second one is a classical representation of the uniform distribution on a simplex~\cite{Devroye1986}. 

\begin{lemma}[The regions of the membrane]
\label{lem:regions}
Let $(p_1,\dots,p_N)$ be the barycentric coordinates of the on-simplex particle $\rpar$, and $(\lambda_1,\dots,\lambda_N)$ those of a point $\bm{\lambda}\neq\rpar$ of the simplex. With the convention $\lambda_j/0=+\infty$, let
\begin{equation}
t(\bm{\lambda})=\min_j\frac{\lambda_j}{p_j}
\label{eq:depth}
\end{equation}
Then:
\begin{itemize}
\item[(a)] for every $i$ such that $p_i>0$, $\bm{\lambda}\in A_i$ if and only if $\lambda_i/p_i=t(\bm{\lambda})$, i.e., if $\lambda_i/p_i\leq\lambda_j/p_j$, for all $j$;
\item[(b)] $0\leq t(\bm{\lambda})<1$, and $\bm{\lambda}=t(\bm{\lambda})\,\rpar+[1-t(\bm{\lambda})]\,\mathbf{f}$, where $\mathbf{f}$ is the point at which the half-line issued from $\rpar$ and passing through $\bm{\lambda}$ leaves the simplex. Hence, for $p_i>0$, $\bm{\lambda}\in A_i$ if and only if this half-line leaves the simplex through the facet $F_i$;
\item[(c)] if $\bm{\lambda}$ is uniformly distributed in the simplex, then $\mu_N\big(t(\bm{\lambda})>u\big)=(1-u)^{N-1}$, for $0\leq u\leq 1$, whatever the position $\rpar$ of the particle.
\end{itemize}
\end{lemma}

\begin{proof}
(a)-(b) A point of $A_i$ is of the form $\bm{\lambda}=t\,\rpar+(1-t)\,\mathbf{f}$, with $0\leq t\leq 1$ and $\mathbf{f}\in F_i$, i.e., with $f_i=0$. For such a point, $\lambda_i=t\,p_i$ and $\lambda_j=t\,p_j+(1-t)f_j\geq t\,p_j$, so that $\lambda_i/p_i=t\leq\lambda_j/p_j$ for all $j$. Conversely, let $t=\lambda_i/p_i$ be the smallest of the ratios. Then $t<1$, since otherwise $\lambda_j\geq t\,p_j\geq p_j$, for all $j$, and the differences $\lambda_j-p_j$, being non-negative and of zero sum, would all vanish, i.e., $\bm{\lambda}=\rpar$. Hence, the numbers $f_j=(\lambda_j-t\,p_j)/(1-t)$ are non-negative, sum to one, and $f_i=0$: they are the barycentric coordinates of a point $\mathbf{f}\in F_i$ such that $\bm{\lambda}=t\,\rpar+(1-t)\,\mathbf{f}$. 

(c) Up to a set of zero volume, $t(\bm{\lambda})>u$ if and only if $\lambda_j>u\,p_j$, for all $j$, i.e., if and only if $\bm{\lambda}=u\,\rpar+(1-u)\,\bm{\lambda}'$, where $\bm{\lambda}'$ is the point of the interior of the simplex with barycentric coordinates $\lambda'_j=(\lambda_j-u\,p_j)/(1-u)$. The points with $t(\bm{\lambda})>u$ thus form the image of the simplex under the homothety of center $\rpar$ and ratio $1-u$ [Fig.~\ref{fig:distances}(b)], whose volume is $(1-u)^{N-1}$ times that of the simplex.
\end{proof}

\noindent The number $t(\bm{\lambda})$, which we call the \emph{relative depth} of $\bm{\lambda}$, is the distance of the disintegration point from the point $\mathbf{f}$ where its half-line reaches the boundary, in units of the distance of the particle from $\mathbf{f}$, since, by (b), $\bm{\lambda}-\mathbf{f}=t(\bm{\lambda})\,(\rpar-\mathbf{f})$. Equivalently, if $\bm{\lambda}\in A_i$, then $\lambda_i=t(\bm{\lambda})\,p_i$, so that, by~(\ref{eq:distance}),
\begin{equation}
d(\bm{\lambda},F_i)=t(\bm{\lambda})\,d(\rpar,F_i)
\label{eq:depth-distance}
\end{equation}
i.e., the relative depth is the ratio between the distances of the disintegration point and of the particle from the facet which determines the outcome. Point (a) shows that the outcome only depends on the ratios between the coordinates of the disintegration point and those of the particle: it is not affected if all the $\lambda_j$, or all the $p_j$, are multiplied by a same positive number. This elementary remark is the key to what follows.

\begin{lemma}[Uniform points of a simplex]
\label{lem:uniform}
Let $E_1,\dots,E_N$ be independent exponential random variables of unit mean, which we write $E_j\sim\Exp(1)$, i.e., such that $\Pp(E_j>s)=e^{-s}$, for $s\geq 0$, and let $S_N=E_1+\dots+E_N$. Then the point of $\triangle_{N-1}$ with barycentric coordinates
\begin{equation}
\lambda_j=\frac{E_j}{S_N},\quad j=1,\dots,N
\label{eq:lambdaE}
\end{equation}
is uniformly distributed in the simplex, and is independent of $S_N$.
\end{lemma}

\begin{proof}
The joint probability density of $(E_1,\dots,E_N)$ is $e^{-(s_1+\dots+s_N)}$ on the set of the points with positive coordinates, and it only depends on their sum $s$. With the variables $(s_1,\dots,s_{N-1},s)$, the change of variables having unit Jacobian, the joint density of $(E_1,\dots,E_{N-1},S_N)$ is $e^{-s}$, on the set $\{s_j\geq 0,\ s_1+\dots+s_{N-1}\leq s\}$, whose volume is $s^{N-1}/(N-1)!$. For a fixed $s$, this density does not depend on $(s_1,\dots,s_{N-1})$: conditionally on $S_N=s$, the point $(E_1,\dots,E_N)$ is uniformly distributed on the set $\{s_j\geq 0,\ \sum_js_j=s\}$, which is the image of $\{\lambda_j\geq 0,\ \sum_j\lambda_j=1\}$ under the multiplication by $s$. This map preserving uniformity, the $E_j/S_N$ are uniformly distributed on the latter set, whatever the value of $s$, hence independently of $S_N$, and so is the point~(\ref{eq:barycentric}), the correspondence being affine.
\end{proof}

\section{The unit ball is too small}
\label{sec:unit}

The volume of a ball of radius $R$ in $\R^n$, and that of a regular $(N-1)$-dimensional simplex $R\,\triangle_{N-1}=\{R\bm{\lambda}:\bm{\lambda}\in\triangle_{N-1}\}$ inscribed in a sphere of radius $R$, are
\begin{equation}
\vol\big[B_R(\R^n)\big]=\frac{\pi^{n/2}}{\Gamma(\frac{n}{2}+1)}\,R^n,\quad \vol\big[R\,\triangle_{N-1}\big]=\frac{\sqrt{N}}{(N-1)!}\left(\frac{N}{N-1}\right)^{\frac{N-1}{2}}\!\!\!\! R^{N-1}
\label{eq:vols}
\end{equation}
the second formula following from the volume $\frac{a^n}{n!}\sqrt{\frac{n+1}{2^n}}$ of a regular $n$-dimensional simplex of edge $a$, with $n=N-1$ and $a=R\sqrt{2N/(N-1)}$. The $n$-th root of the volume of an $n$-dimensional body, which we call its \emph{volume per dimension}, is the side of the cube having the same volume. By Stirling's formula, $\Gamma(m+1)^{1/m}\sim m/e$, so that, for the generalized Bloch ball ($n=N^2-1$) and the measurement simplex ($n=N-1$),
\begin{equation}
\vol\big[B_R(\R^{N^2-1})\big]^{\frac{1}{N^2-1}}\sim\sqrt{2\pi e}\ \frac{R}{N},\quad \vol\big[R\,\triangle_{N-1}\big]^{\frac{1}{N-1}}\sim e\ \frac{R}{N}
\label{eq:volperdim}
\end{equation}
For $R=1$, both vanish as $N\to\infty$, and the Born probabilities~(\ref{eq:volumes}) appear in the limit as undefined ``zero over zero'' ratios~\cite{AS2017b}, although they are well defined for every finite $N$. To understand what the problem is, we have to look at what happens, in the limit, to the states and to the measurement interactions.

\subsection{Limit of states}
\label{subsec:states}

Let $\mathcal{H}$ be infinite-dimensional and separable, $A=\sum_{i=1}^\infty a_iP_{a_i}$ an observable with a non-degenerate pure point spectrum, $Q_N=\sum_{i\leq N}P_{a_i}$, and $D$ a state. The compressed states $D_N=Q_NDQ_N/\Tr DQ_N$ (defined for $N$ large enough, so that $\Tr DQ_N>0$), of $N$-dimensional entities with Hilbert spaces $Q_N\mathcal{H}$, converge to $D$~\cite{AS2014}, and by~(\ref{eq:state}) the vector $\mathbf{r}_N$ representing $D_N$ is such that
\begin{equation}
\mathbf{r}_N\cdot\bm{\Lambda}=\sqrt{\frac{2N}{N-1}}\,\Big(D_N-\frac{Q_N}{N}\Big)
\label{eq:r-operator}
\end{equation}
The scalar product of the generalized Bloch space being the Hilbert-Schmidt one, $\mathbf{r}\cdot\mathbf{s}=\frac{1}{2}\Tr(\mathbf{r}\cdot\bm{\Lambda})(\mathbf{s}\cdot\bm{\Lambda})$ (Section~\ref{sec:EBR}), for which the norm of $Q_N/N$ is $1/\sqrt{2N}\to 0$, these operators converge to $\sqrt{2}\,D$: the limit of the generalized Bloch space is the real Hilbert space $\mathcal{B}_\infty$ of the self-adjoint Hilbert-Schmidt operators on $\mathcal{H}$, with the scalar product $(X|Y)=\frac{1}{2}\Tr XY$, which we call the \emph{limit Bloch space}, and in which $D$ is represented by $\sqrt{2}\,D$. As described in~\cite{AS2017b}, the vertices $\mathbf{n}_i=\sqrt{2}\,P_{a_i}$ become orthonormal, the Born rule becomes $p_i=(\sqrt{2}D|\sqrt{2}P_{a_i})=(\mathbf{r}|\mathbf{n}_i)$, the simplex becomes $\triangle_\infty=\{\sum_is_i\mathbf{n}_i:\ s_i\geq 0,\ \sum_is_i=1\}$, which does not contain the origin, and the states lie in the unit ball, since $\|\sqrt{2}\,D\|^2=\Tr D^2\leq 1$, but have no interior points, since the eigenvalues of a density operator accumulate at zero and an arbitrarily small perturbation can make one of them negative. The unit ball thus has a perfectly good limit as far as the states are concerned.

\subsection{Shrinking membranes}

The situation is different for the measurement interactions. In the unit ball, the vertices of the measurement simplex are at distance $1$ from the center, and its facets at distance $\frac{1}{N-1}$. The following proposition gives the third length scale, that of the disintegration points.

\begin{proposition}[The membrane in the unit ball]
\label{prop:scales}
In the generalized Bloch ball of unit radius:
\begin{itemize}
\item[(a)] if $\bm{\lambda}$ is uniformly distributed in the simplex, its mean square distance from the center is
\begin{equation}
\E\big[\|\bm{\lambda}\|^2\big]=\frac{1}{N+1}
\label{eq:shell}
\end{equation}
and $(N+1)\,\|\bm{\lambda}\|^2\to 1$ in probability, as $N\to\infty$;
\item[(b)] the relative depth~(\ref{eq:depth}) of a uniformly distributed $\bm{\lambda}$ has mean $\E[t(\bm{\lambda})]=\frac{1}{N}$, whatever the state. 
\end{itemize}
\end{proposition}

\begin{proof}
(a) Using~(\ref{eq:barycentric}), $\mathbf{n}_j\cdot\mathbf{n}_k=-\frac{1}{N-1}$, for $j\neq k$, and $\sum_{j\neq k}\lambda_j\lambda_k=1-\sum_j\lambda_j^2$, we obtain
\begin{equation}
\|\bm{\lambda}\|^2=\sum_j\lambda_j^2-\frac{1}{N-1}\sum_{j\neq k}\lambda_j\lambda_k=\frac{N}{N-1}\sum_j\lambda_j^2-\frac{1}{N-1}
\label{eq:norm}
\end{equation}
The coordinate $\lambda_1$ has the probability density $(N-1)(1-\lambda)^{N-2}$ on $[0,1]$ (see the proof of Theorem~\ref{thm:limit} below, with $k=1$), so that $\E[\lambda_1^2]=\frac{2}{N(N+1)}$, and by symmetry $\E[\sum_j\lambda_j^2]=\frac{2}{N+1}$. Using (\ref{eq:norm}), this 
gives~(\ref{eq:shell}). Moreover, the $E_j$ being of unit mean, and $\E[E_1^2]=2$, by Lemma~\ref{lem:uniform} and the law of large numbers, $N\sum_j\lambda_j^2=\frac{1}{N}\sum_jE_j^2\big/(S_N/N)^2\to\E[E_1^2]/\E[E_1]^2=2$, so that $(N+1)\,\|\bm{\lambda}\|^2=\frac{N+1}{N-1}\big(N\sum_j\lambda_j^2-1\big)\to 1$. 

(b) $\E[t(\bm{\lambda})]=\int_0^1\Pp\big(t(\bm{\lambda})>u\big)\,du=\int_0^1(1-u)^{N-1}du=\frac{1}{N}$, by Lemma~\ref{lem:regions}(c).
\end{proof}

\noindent Point (a) is an instance of the concentration of measure in high dimension. By Markov's inequality, $\Pp(\|\bm{\lambda}\|>\delta)\leq\frac{1}{(N+1)\delta^2}$, so that, in the limit, the uniform membrane becomes a Dirac measure concentrated at the origin, which is not even a point of $\triangle_\infty$, and cannot distribute weights among different regions: the measurement interactions, and with them the mechanism producing the probabilities, have disappeared from the picture.

The reason is that the particle and the disintegration point live in the same simplex, but at different scales. For a fixed state, the finitely many outcomes which carry almost all the probability do not depend on $N$ (up to the normalization of the compressed states $D_N$), and by~(\ref{eq:distance}) the particle remains at distances of order one from the corresponding facets. On the contrary, the barycentric coordinates of a uniformly distributed point are typically all of order $1/N$, and so is, by point (b), its relative depth, whatever the state: to see the measurement interactions, the picture must be magnified by a factor of order $N$. Note that a typical disintegration point is thus at a distance of order $1/N$ from all the facets, like the center, but, by point (a), at a distance of order $1/\sqrt{N}$ from the center, well outside the inscribed sphere: in high dimension, a simplex is a ``spiky'' object.

\section{The renormalized EBR}
\label{sec:renormalized}

Nothing in the EBR forces the radius of the Bloch ball to be equal to one: multiplying all the vectors by a same factor $R$ only changes the unit of length, and all the ratios of volumes, hence the probabilities~(\ref{eq:volumes}), remain the same. In a ball of radius $R$, the inscribed sphere has radius $R/(N-1)$, the height of the simplex is $RN/(N-1)$ and, by~(\ref{eq:distance}), the distance of $\bm{\lambda}$ from the facet $F_j$ is $\lambda_jRN/(N-1)$. Proposition~\ref{prop:scales} then suggests the following choice.
\begin{definition}[Renormalized EBR]
\label{def:renormalized}
The \emph{renormalized} extended Bloch representation (REBR) is the EBR in which the generalized Bloch ball has radius
\begin{equation}
R_N=N-1
\end{equation}
i.e., in which the state $D(\mathbf{r})$ is represented by the vector $\tilde{\mathbf{r}}=(N-1)\,\mathbf{r}$.
\end{definition}

\noindent In the REBR, the sphere inscribed in the measurement simplex has \emph{unit radius}, whatever $N$, the height of the simplex is $N$, and the vertices are at distance $N-1$ from the center. The distances of the point $\bm{\lambda}$ of the simplex and of the on-simplex particle from the facet $F_j$ are
\begin{equation}
\xi_j\equiv d(\bm{\lambda},F_j)=N\lambda_j,\quad\eta_j\equiv d(\rpar,F_j)=Np_j
\label{eq:xi}
\end{equation}
The $\xi_j$ are non-negative numbers of sum $N$, i.e., of unit mean, all equal to one for the center, whereas the $i$-th vertex is at distance $N$ from $F_i$, and zero from the other facets. By Lemma~\ref{lem:regions}(a), and since $\xi_j/\eta_j=\lambda_j/p_j$, the disintegration point belongs to the region $A_i$, and the outcome is $a_i$, if and only if
\begin{equation}
\frac{\xi_i}{\eta_i}\leq\frac{\xi_j}{\eta_j},\quad\text{for all }j
\label{eq:rule}
\end{equation}

\subsection{Renormalized operators}

Expressing~(\ref{eq:state}) by means of the vector $\tilde{\mathbf{r}}=(N-1)\,\mathbf{r}$, one obtains
\begin{equation}
\tilde{D}(\tilde{\mathbf{r}})\equiv N\,D(\mathbf{r})=\mathbb{I}+\tilde{c}_N\,\tilde{\mathbf{r}}\cdot\bm{\Lambda},\quad \tilde{c}_N=\sqrt{\frac{N}{2(N-1)}}\label{eq:X}
\end{equation}
The renormalized vector thus represents the operator-state \emph{normalized to the dimension}, $\Tr\tilde{D}=N$, whose eigenvalues have unit mean. The center of the ball represents the identity operator, the $i$-th vertex the operator $NP_{a_i}$ and, since~(\ref{eq:state}) is affine, the point $\bm{\lambda}=\sum_j\lambda_j\mathbf{n}_j$ of the simplex represents $N\sum_j\lambda_jD(\mathbf{n}_j)$, i.e.,
\begin{equation}
\tilde{D}(\bm{\xi})=\sum_{j=1}^N\xi_jP_{a_j}
\label{eq:Xxi}
\end{equation}
the distances from the facets being the eigenvalues of the renormalized operator.

This is the same kind of renormalization as that of the eigenvectors of a continuous observable. For a particle on a lattice of $N$ sites, of spacing $\epsilon$, in a box of length $L=N\epsilon$, the continuum limit is taken by means of the unnormalized vectors $|x_j\rangle=|v_j\rangle/\sqrt{\epsilon}$, where the $|v_j\rangle$ are the normalized position eigenvectors, whose squared norm $1/\epsilon$ diverges, but whose scalar products with a vector-state have squared moduli $p_j/\epsilon$, which converge to the Born density (Section~\ref{sec:continuous}). With $|\tilde{a}_j\rangle=|a_j\rangle/\sqrt{\epsilon}$, the vertices of the simplex represent the operators
\begin{equation}
NP_{a_j}=N|a_j\rangle\langle a_j|=N\epsilon\,|\tilde{a}_j\rangle\langle\tilde{a}_j|=L\,|\tilde{a}_j\rangle\langle\tilde{a}_j|
\label{eq:vertex-x}
\end{equation}
i.e., up to the constant $L$, the ``projections'' onto the unnormalized eigenvectors: the growth of the radius of the ball, proportional to $N=L/\epsilon$, is that of the squared norm of these vectors.

\subsection{Non-shrinking volumes}

If the radius of the generalized Bloch ball is $R_N=\kappa N$, or $R_N=\kappa(N-1)$, with $\kappa>0$, then, according to~(\ref{eq:volperdim}), we have the limits:
\begin{equation}
\lim_{N\to\infty}\vol\big[B_{R_N}(\R^{N^2-1})\big]^{\frac{1}{N^2-1}}=\kappa\sqrt{2\pi e},\quad \lim_{N\to\infty}\vol\big[R_N\,\triangle_{N-1}\big]^{\frac{1}{N-1}}=\kappa\,e
\end{equation}

\noindent A radius proportional to $N$ thus cures the two collapses at once: that of the ball, which has $N^2-1$ dimensions, and that of the simplex, which has only $N-1$. This coincidence could not be taken for granted, and it is essential: had the two bodies required different growths of the radius, no renormalization of the ball could have preserved both.

For $\kappa=1$, using~(\ref{eq:vols}) and Stirling's formula, $\vol\big[(N-1)\triangle_{N-1}\big]\sim\sqrt{e/2\pi}\ e^{N-1}$, whereas for $R_N=(N-1)/e$ the volume of the simplex tends to $\sqrt{e/2\pi}$, and~(\ref{eq:volumes}) becomes, in the limit, a ratio of two finite numbers. Since only the ratios of volumes matter, there is however no need for such a fine-tuning, and we keep the choice of Definition~\ref{def:renormalized}.

\section{The limit of the uniform membrane}
\label{sec:limit}

In this section we show that, in the renormalized ball, the distances of the measurement interactions from the facets become, as $N\to\infty$, independent exponential variables, and that the renormalization of Definition~\ref{def:renormalized} is essentially the only one leading to a non-degenerate limit.

\subsection{Distances as independent exponential variables}

\begin{theorem}[Limit of the uniform membrane]
\label{thm:limit}
Let $\bm{\lambda}^{(N)}$ be uniformly distributed in $\triangle_{N-1}$, and let $\xi^{(N)}_j=N\lambda^{(N)}_j$, $j=1,\dots,N$, be its distances from the facets in the renormalized ball. Then, for every fixed $k$, the joint distribution of $(\xi^{(N)}_1,\dots,\xi^{(N)}_k)$ converges, as $N\to\infty$, to that of $k$ independent $\Exp(1)$ random variables. More precisely:
\begin{itemize}
\item[(a)] for $N>k$, the probability density of $(\xi^{(N)}_1,\dots,\xi^{(N)}_k)$ is
\begin{equation}
f_k^{(N)}(\xi_1,\dots,\xi_k)=\frac{(N-1)!}{(N-k-1)!\,N^k}\left(1-\frac{\xi_1+\dots+\xi_k}{N}\right)^{N-k-1}
\label{eq:density}
\end{equation}
on the set $\{\xi_j\geq 0,\ \xi_1+\dots+\xi_k\leq N\}$, and it converges to $e^{-(\xi_1+\dots+\xi_k)}$, pointwise and in $L^1$;
\item[(b)] if $E_1,E_2,\dots$ is a sequence of independent $\Exp(1)$ random variables, and $\bm{\lambda}^{(N)}$ is constructed as in Lemma~\ref{lem:uniform} from its first $N$ terms, then $\xi^{(N)}_j\to E_j$ almost surely, for every $j$.
\end{itemize}
\end{theorem}

\begin{proof}
(a) The first $N-1$ barycentric coordinates are uniformly distributed on the set $\{\lambda_j\geq 0,\ \sum_{j<N}\lambda_j\leq 1\}$, of volume $1/(N-1)!$ (proof of Lemma~\ref{lem:uniform}), i.e., with density $(N-1)!$. Integrating over $\lambda_{k+1},\dots,\lambda_{N-1}$, which range in a set of the same kind, of dimension $N-k-1$ and volume $(1-s)^{N-k-1}/(N-k-1)!$, with $s=\lambda_1+\dots+\lambda_k$, we find that the density of $(\lambda_1,\dots,\lambda_k)$ is $\frac{(N-1)!}{(N-k-1)!}(1-s)^{N-k-1}$, and the change of variables $\xi_j=N\lambda_j$, of Jacobian $N^{-k}$, gives~(\ref{eq:density}). As $N\to\infty$, the prefactor $\prod_{m=1}^k\big(1-\frac{m}{N}\big)$ tends to one and $(1-\frac{s}{N})^{N-k-1}\to e^{-s}$. Since the $f_k^{(N)}$ and their limit are probability densities, the pointwise convergence implies the convergence in $L^1$ (Scheffé's lemma~\cite{Billingsley1995}), i.e., the uniform convergence of the probabilities of all events. (b) By~(\ref{eq:lambdaE}), $\xi^{(N)}_j=E_j/(S_N/N)$, and $S_N/N\to 1$ almost surely, by the strong law of large numbers.
\end{proof}

\noindent By symmetry, the statement holds for any $k$ facets: seen from the renormalized ball, the disintegration point is at independent exponential distances of unit mean from any finite number of facets, the constraint $\sum_j\xi_j=N$ becoming ineffective in the limit. This is the simplex version of the classical fact that the coordinates of a point uniformly distributed on the sphere of radius $\sqrt{N}$ of $\R^N$ become independent standard Gaussian variables~\cite{DiaconisFreedman1987}, and the equivalence of the microcanonical and canonical ensembles of statistical mechanics~\cite{Khinchin1949}, the $\xi_j$ being the energies of $N$ subsystems of total energy $N$.

\subsection{Uniqueness of renormalization}

Is the growth $R_N\propto N$ the only possible one, and could a cleverer choice keep the membrane inside a ball of finite radius? The following proposition answers the first question.

\begin{proposition}[Other radii]
\label{prop:radii}
Let $R_N$ be an arbitrary sequence of radii, and let $\bm{\rho}^{(N)}=R_N\bm{\lambda}^{(N)}$ be the disintegration point of the uniform membrane in the ball of radius $R_N$. Its distance $d(\bm{\rho}^{(N)},F_j)$ from the facet $F_j$, its component along the direction of the $j$-th vertex, and its mean square distance from the center are, respectively, 
\begin{equation}
\frac{R_N}{N-1}\xi^{(N)}_j,\quad\bm{\rho}^{(N)}\! \cdot\mathbf{n}_j=\frac{R_N}{N-1}\big(\xi^{(N)}_j-1\big),\quad\E\left[\|\bm{\rho}^{(N)}\|^2\right]=\frac{R_N^2}{N+1}
\label{eq:components}
\end{equation}
Consequently, as $N\to\infty$:
\begin{itemize}
\item[(a)] if $R_N/N\to 0$, the first two quantities tend to zero in probability, for every $j$;
\item[(b)] if $R_N/N\to\kappa\in(0,\infty)$, they converge in distribution to $\kappa E_j$ and $\kappa(E_j-1)$, with $E_j\sim\Exp(1)$, and $\|\bm{\rho}^{(N)}\|\to\infty$ in probability;
\item[(c)] if $R_N/N\to\infty$, their absolute values diverge in probability;
\item[(d)] the mean square distance from the center remains bounded only if $R_N$ is at most of order $\sqrt{N}$, which is case (a).
\end{itemize}
\end{proposition}

\begin{proof}
The first expression is~(\ref{eq:distance}) multiplied by $R_N$, and the third is~(\ref{eq:shell}) multiplied by $R_N^2$. For the second, $\bm{\rho}^{(N)}\cdot\mathbf{n}_j=R_N\sum_k\lambda_k\,\mathbf{n}_k\cdot\mathbf{n}_j=R_N\big(\lambda_j-\frac{1-\lambda_j}{N-1}\big)=\frac{R_N}{N-1}(N\lambda_j-1)$. 
Points (a)-(c) then follow from Theorem~\ref{thm:limit}: the random variables $\xi^{(N)}_j$ converge in distribution to $E_j$, and they are multiplied by the deterministic factor $R_N/(N-1)$, which tends to zero, to $\kappa$, or to infinity [for (c), note that the limits $E_j$ and $E_j-1$ vanish with probability zero].
Point (d), as well as the last statement of (b), follows from~(\ref{eq:components}) and Proposition~\ref{prop:scales}(a): $R_N^2/(N+1)$ remains bounded if and only if $R_N$ is at most of order $\sqrt{N}$, in which case $R_N/N\to 0$.
\end{proof}

\noindent Hence, up to the irrelevant constant $\kappa$, the renormalization of Definition~\ref{def:renormalized} is the only growth of the radius for which the membrane has a non-degenerate limit. But the proposition also shows that this limit has a price: the coordinates of the disintegration point converge, but its distance from the center, of order $\sqrt{N}$, diverges. Conversely, when the distance from the center is kept finite, the coordinates vanish. 

Could a cleverer choice, not necessarily a rescaling of the ball, keep the limit membrane at a finite distance from the center? The answer is negative, because of a symmetry: the joint distribution of $(\xi_1,\dots,\xi_N)$ is invariant under the permutations of the coordinates. Before the particle has fallen onto it, the membrane does not distinguish between the outcomes, whose labeling is arbitrary, and any procedure for taking the limit that treats all the outcomes on the same footing preserves this invariance, for the permutations of finitely many indices (\emph{exchangeability}). The following proposition shows that such a symmetry is incompatible with a finite norm.

\begin{proposition}[No limit membrane at a finite distance from the center]
\label{prop:nogo}
Let $y_1,y_2,\dots$ be a sequence of real random variables whose joint probability distribution is invariant under the permutations of any finite number of indices. If $\sum_jy_j^2<\infty$ with probability one, then $y_j=0$ for all $j$, with probability one.
\end{proposition}

\begin{proof}
Intuitively, the invariance under permutations means that all the coordinates are ``equally large,'' so that they cannot all be the terms of a convergent series, unless they are all zero. More precisely, if the series converges, its general term tends to zero: $y_j\to 0$, as $j\to\infty$, with probability one, so that the distribution of $y_j$ tends to the Dirac measure concentrated at zero. But, by the invariance under permutations, all the $y_j$ have the same distribution, which must therefore coincide with its limit: $y_j=0$ with probability one, for every $j$.
\end{proof}

\noindent In $\mathcal{B}_\infty$, the vertex directions $\mathbf{n}_j$ are orthonormal, so that, by Proposition~\ref{prop:nogo}, the only  
exchangeable limit membrane whose disintegration points are almost surely vectors of $\mathcal{B}_\infty$ is the trivial one, found in Section~\ref{sec:unit}: a non-trivial limit membrane has infinitely many components of order one, hence an infinite norm. Like the Gaussian measure of Wiener and Lévy, viewed as the uniform distribution on a sphere of infinite dimension and infinite radius~\cite{McKean1973}, it lives on a simplex of infinite height, in a generalized Bloch ball of infinite radius, its points being described by the sequences of their distances from the facets.

\section{The infinite membrane}
\label{sec:infinite-membrane}

In this section we define the limit object, which we call the infinite membrane, describe its main properties, and show that it is the projective limit of the finite-dimensional uniform membranes.

\subsection{Definition and main properties}

\begin{definition}[Infinite membrane]
\label{def:membrane}
Let $\Rinf=[0,\infty)^{\N}$ be the set of the sequences $\bm{\xi}=(\xi_1,\xi_2,\dots)$ of non-negative numbers, and $\mu_\infty$ the probability measure on $\Rinf$ under which the $\xi_j$ are independent $\Exp(1)$ random variables. The probability space $(\Rinf,\mu_\infty)$ will be called the \emph{infinite membrane}. As for the finite membranes, each point $\bm{\xi}$ of the infinite membrane corresponds to a ``hidden'' \emph{measurement interaction}, and $\xi_j$ is its distance from the $j$-th facet.
\end{definition}

\noindent We recall that an infinite product of probability measures is always a well-defined probability measure (Kolmogorov's extension theorem~\cite{Billingsley1995}): the events are generated by the conditions involving a finite number of coordinates, whose probabilities are given by the corresponding finite products. Theorem~\ref{thm:limit} says that $\mu_\infty$ is the limit of the uniform distributions $\mu_N$ of the membranes with $N$ outcomes, described by means of the distances from the facets in the renormalized balls.

\begin{proposition}[Where the infinite membrane lives]
\label{prop:where}
For $\mu_\infty$-almost every $\bm{\xi}$:
\begin{itemize}
\item[(a)] $\xi_j>0$, for all $j$;
\item[(b)] $\displaystyle\lim_{n\to\infty}\frac{1}{n}\sum_{j=1}^n\xi_j=1$;
\item[(c)] $\displaystyle\lim_{n\to\infty}\frac{1}{n}\sum_{j=1}^n(\xi_j-1)^2=1$, so that $\sum_j(\xi_j-1)^2=\infty$;
\item[(d)] $\inf_j\xi_j=0$ and $\sup_j\xi_j=\infty$.
\end{itemize}
\end{proposition}

\begin{proof}
(a) The countable union of the events $\{\xi_j=0\}$, each of zero probability, has zero probability. (b), (c) The $\xi_j$, and the $(\xi_j-1)^2$, are independent and identically distributed, with means $1$ and $\E[\xi_j^2]-1=1$, and the strong law of large numbers applies~\cite{Billingsley1995}. (d) For $c>0$, the probabilities that the first $n$ distances are all larger than $c$, or all smaller than or equal to $c$, are $e^{-cn}$ and $(1-e^{-c})^n$, which tend to zero as $n\to\infty$. Hence, with probability one, there are distances smaller than or equal to $c$, and distances larger than $c$, simultaneously for the countably many values $c=\frac{1}{k}$ and $c=k$, $k=1,2,\dots$, so that $\inf_j\xi_j=0$ and $\sup_j\xi_j=\infty$.
\end{proof}

\noindent By (b), the measurement interactions of the infinite membrane are sequences of \emph{unit mean}: the infinite membrane is a simplex whose inscribed sphere still has unit radius, but whose height is infinite. By Proposition~\ref{prop:radii}, the $\xi_j-1$ are the components of the measurement interaction along the vertex directions, so that, by (c), the measurement interaction lies infinitely far from the center, outside $\mathcal{B}_\infty$, and by (d) it comes arbitrarily close to infinitely many facets. In terms of operators, by~(\ref{eq:Xxi}), $\bm{\xi}$ corresponds to the positive, unbounded operator $\tilde{D}(\bm{\xi})=\sum_{j}\xi_jP_{a_j}$, of infinite trace, whose quadratic form $\sum_j\xi_j|\langle a_j|\psi\rangle|^2$ has mean $\langle\psi|\psi\rangle$, hence is almost surely finite for every fixed $|\psi\rangle$.

\subsection{The infinite membrane as a projective limit}

\begin{figure}[tp]
\centering
\includegraphics[width=0.8\textwidth]{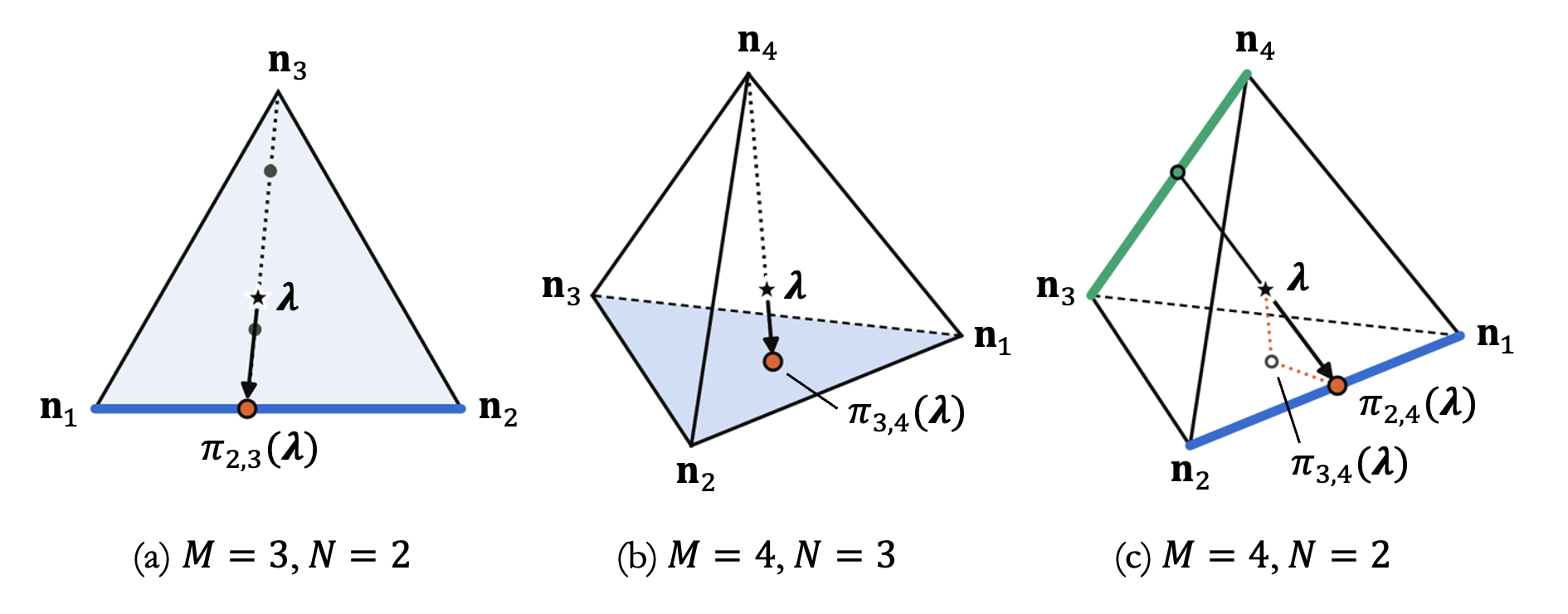} 
\caption{The shadow maps $\pi_{N,M}$, from the simplex with $M$ vertices to the face generated by its first $N$ vertices (thick line or shaded triangle). (a) $M=3$, $N=2$: $\pi_{2,3}(\bm{\lambda})$ is the point where the half-line issued from $\mathbf{n}_3$ and passing through $\bm{\lambda}$ (dotted) reaches the opposite edge, as if $\bm{\lambda}$ were illuminated by a point source of light located at $\mathbf{n}_3$. All the points of this half-line, such as the two small grey dots, have the same shadow. (b) $M=4$, $N=3$: the same construction, from the vertex $\mathbf{n}_4$. (c) $M=4$, $N=2$: $\pi_{2,4}(\bm{\lambda})$ is the endpoint, on the edge joining $\mathbf{n}_1$ and $\mathbf{n}_2$, of the segment that joins the two opposite edges and passes through $\bm{\lambda}$. The same point is obtained in two steps (dotted path), in accordance with $\pi_{2,3}\circ\pi_{3,4}=\pi_{2,4}$; see Theorem~\ref{thm:projective}(i).}
\label{fig:shadow}
\end{figure}

The infinite membrane contains all the finite-dimensional membranes, as its ``shadows''. For $\bm{\xi}\in\Rinf$ such that $\xi_1+\dots+\xi_N>0$, let $\pi_{N,\infty}(\bm{\xi})$ be the point of $\triangle_{N-1}$ with barycentric coordinates
\begin{equation}
[\pi_{N,\infty}(\bm{\xi})]_j=\frac{\xi_j}{\xi_1+\dots+\xi_N},\quad j=1,\dots,N
\label{eq:shadow}
\end{equation}
which we call the $N$-dimensional \emph{shadow} of $\bm{\xi}$, and, for $N<M$, let $\pi_{N,M}$ map the point of $\triangle_{M-1}$ with barycentric coordinates $(\lambda_1,\dots,\lambda_M)$ to the point of $\triangle_{N-1}$ with barycentric coordinates $\lambda_j/(\lambda_1+\dots+\lambda_N)$, $j\leq N$. Identifying $\triangle_{N-1}$ with the face of $\triangle_{M-1}$ generated by its first $N$ vertices, we have $\bm{\lambda}=s\,\pi_{N,M}(\bm{\lambda})+(1-s)\,\mathbf{g}$, with $s=\lambda_1+\dots+\lambda_N$ and $\mathbf{g}$ a point of the opposite face: $\pi_{N,M}(\bm{\lambda})$ is the endpoint, on the first face, of the segment joining the two faces through $\bm{\lambda}$. For $M=N+1$, it is the shadow of $\bm{\lambda}$ cast on the facet $F_M$ by a point source of light located at the vertex $\mathbf{n}_M$, hence the name \emph{shadow map} (Fig.~\ref{fig:shadow}).

Physically, if the state is such that only the first $N$ outcomes have a non-zero probability, the on-simplex particle lies on that face, and by Lemma~\ref{lem:regions}(a) the outcome of the measurement only depends on $\pi_{N,M}(\bm{\lambda})$: the measurement with $M$ outcomes reduces to one with $N$ outcomes, whose measurement interaction is $\pi_{N,M}(\bm{\lambda})$.

We say that a subset $C$ of $\Rinf$ is a \emph{cone} if $\bm{\xi}\in C$ implies $c\,\bm{\xi}\in C$, for all $c>0$, i.e., if it is a union of half-lines (rays) issued from the origin.

\begin{theorem}[Projective limit]
\label{thm:projective}
\begin{itemize}
\item[]
\item[(i)] The shadow maps are compatible: $\pi_{N,M}\circ\pi_{M,K}=\pi_{N,K}$, for $N<M<K$. In particular, for $K=\infty$, $\pi_{N,M}\circ\pi_{M,\infty}=\pi_{N,\infty}$.
\item[(ii)] The shadow map $\pi_{N,M}$ transforms the uniform distribution $\mu_M$ of $\triangle_{M-1}$ into the uniform distribution $\mu_N$ of $\triangle_{N-1}$, and the same holds for $M=\infty$, with the uniform distribution $\mu_M$ replaced by the probability law $\mu_\infty$ of the infinite membrane: the shadows of the infinite membrane are the uniform membranes of all dimensions.
\item[(iii)] For $\mu_\infty$-almost every $\bm{\xi}$, the distances from the facets of the shadows of $\bm{\xi}$, in the renormalized balls, converge to those of $\bm{\xi}$: $N\,[\pi_{N,\infty}(\bm{\xi})]_j\to\xi_j$, as $N\to\infty$, for every $j$.
\item[(iv)] Let $\nu$ be a probability measure on $\Rinf$, with $\nu(\xi_j>0,\text{ for all }j)=1$, whose shadows are all uniform, i.e., which is transformed into $\mu_N$ by $\pi_{N,\infty}$, for every $N$. Then $\nu(C)=\mu_\infty(C)$ for every measurable cone $C$. If, moreover, $\frac{1}{n}\sum_{j\leq n}\xi_j\to 1$, for $\nu$-almost every $\bm{\xi}$, then $\nu=\mu_\infty$.
\end{itemize}
\end{theorem}

\begin{proof}
(i) Both sides retain the first $N$ coordinates and divide them by their sum, the intermediate division by a common factor cancelling out. (ii) If $\bm{\xi}$ is distributed according to $\mu_\infty$, its first $N$ coordinates are independent $\Exp(1)$ variables, so that, by Lemma~\ref{lem:uniform}, $\pi_{N,\infty}(\bm{\xi})$ is uniformly distributed in $\triangle_{N-1}$. For finite $M$, $\pi_{M,\infty}(\bm{\xi})$ is uniformly distributed in $\triangle_{M-1}$ and, by (i), its image $\pi_{N,M}[\pi_{M,\infty}(\bm{\xi})]=\pi_{N,\infty}(\bm{\xi})$ is uniformly distributed in $\triangle_{N-1}$. Since the law of the image only depends on that of the point, $\pi_{N,M}$ transforms $\mu_M$ into $\mu_N$.

(iii) $N[\pi_{N,\infty}(\bm{\xi})]_j=\xi_j/\big(\frac{1}{N}\sum_{k\leq N}\xi_k\big)$, and the denominator tends to one by Proposition~\ref{prop:where}(b). 

(iv) By hypothesis, and by Proposition~\ref{prop:where}(a) for $\mu_\infty$, we have $\xi_j>0$ for all $j$, both $\nu$-almost surely and $\mu_\infty$-almost surely, so that the ratios $\varrho_j=\xi_j/\xi_1$, $j\geq 2$, are well defined. For every $N$, $(\varrho_2,\dots,\varrho_N)$ is a function of the shadow $\pi_{N,\infty}(\bm{\xi})$, since $\varrho_j=[\pi_{N,\infty}(\bm{\xi})]_j/[\pi_{N,\infty}(\bm{\xi})]_1$, so that it has the same distribution under $\nu$ and under $\mu_\infty$, and therefore, by Kolmogorov's theorem, so has the whole sequence $(\varrho_j)_{j\geq 2}$. If $C$ is a measurable cone, the condition $\bm{\xi}\in C$ is equivalent (taking $c=1/\xi_1$) to $(1,\varrho_2,\varrho_3,\dots)\in C$, which only involves the ratios. Hence, $\nu(C)=\mu_\infty(C)$. For the last statement, let $\theta(\bm{\xi})=\lim_{n\to\infty}\frac{1}{n}\sum_{j\leq n}\xi_j$. The set on which this limit exists and is positive and finite is a measurable cone of $\mu_\infty$-measure one, hence of $\nu$-measure one, on which the map $\Phi(\bm{\xi})=\bm{\xi}/\theta(\bm{\xi})$ is constant along the rays, so that, for every measurable set $B$, $\{\Phi\in B\}$ is a measurable cone, and $\nu(\Phi\in B)=\mu_\infty(\Phi\in B)$. But $\theta=1$, i.e., $\Phi(\bm{\xi})=\bm{\xi}$, almost surely under both $\mu_\infty$ [Proposition~\ref{prop:where}(b)] and $\nu$ (by hypothesis). Hence $\nu(B)=\mu_\infty(B)$.
\end{proof}

\noindent The uniform membranes of all dimensions thus form a consistent (projective) family, and the infinite membrane, of which they are all shadows, is unique as far as the rays are concerned, which is all that matters (Lemma~\ref{lem:regions}), the point on each ray being fixed by the renormalization. Note that the exponential law is not a hypothesis in (iv), but a consequence of the uniformity of the finite-dimensional membranes.

\section{Measurements with a pure point spectrum}
\label{sec:purepoint}

We now go back to the situation of Section~\ref{sec:unit}, with a state $D$ whose Born probabilities $p_i=\Tr DP_{a_i}$, infinitely many of which can be non-zero, form the sequence $\mathbf{p}=(p_1,p_2,\dots)$ of the barycentric coordinates of the on-simplex particle. It is in this form, like the measurement interactions $\bm{\xi}$, that the particle will be compared with the latter since, in the renormalized ball, only its direction survives in the limit.

\subsection{The regions become cones}

In the renormalized ball, by~(\ref{eq:rule}), the region $A_i$ of the membrane is described by the inequalities $\xi_i/p_i\leq\xi_j/p_j$, which only involve the ratios of the distances from the facets, and still make sense for an infinite sequence. We thus define, for $p_i>0$,
\begin{equation}
C_i(\mathbf{p})=\Big\{\bm{\xi}\in\Rinf:\ \frac{\xi_i}{p_i}<\frac{\xi_j}{p_j}, \text{ for all }j\neq i\Big\}
\label{eq:cone}
\end{equation}
with the convention $\xi_j/0=+\infty$, and $C_i(\mathbf{p})=\emptyset$ if $p_i=0$. The inequalities are taken strict so that the sets are disjoint (the points for which two ratios are equal will be seen to have probability zero).

\begin{theorem}[The Born rule on the infinite membrane]
\label{thm:born}
\begin{itemize}
\item[]
\item[(i)] The sets $C_i(\mathbf{p})$ are cones. If only the first $N$ outcomes have a non-zero probability,  
and $\xi_1+\dots+\xi_N>0$ (so that the shadow $\pi_{N,\infty}(\bm{\xi})$ is defined), $\bm{\xi}\in C_i(\mathbf{p})$ if and only if the barycentric coordinates $\lambda_j$ of the shadow $\pi_{N,\infty}(\bm{\xi})$ satisfy $\lambda_i/p_i<\lambda_j/p_j$, for all $j\leq N$, $j\neq i$: up to a set of zero probability, the cone $C_i(\mathbf{p})$ is the set of the measurement interactions whose shadow lies in the region $A_i$ of $\triangle_{N-1}$.
\item[(ii)] $\mu_\infty$-almost every measurement interaction $\bm{\xi}$ belongs to one and only one of the cones. In other words, the outcome of the measurement is well defined.
\item[(iii)] $\mu_\infty\big[C_i(\mathbf{p})\big]=p_i$, for every $i$.
\item[(iv)] The quantity $T(\bm{\xi})=\min_j\xi_j/p_j$ has the distribution $\Exp(1)$, whatever the state, and it is independent of the outcome: $\mu_\infty\big(C_i(\mathbf{p})\cap\{T>u\}\big)=p_i\,e^{-u}$.
\item[(v)] Let $D_N=Q_NDQ_N/\Tr DQ_N$ be the compressed states of Section~\ref{sec:unit}, and $A^{(N)}_i$ the corresponding regions of $\triangle_{N-1}$. If $\bm{\xi}\in C_i(\mathbf{p})$  
and $\xi_j>0$ for all $j$ (which holds for $\mu_\infty$-almost every $\bm{\xi}$), then $\pi_{N,\infty}(\bm{\xi})\in A^{(N)}_i$, for all $N\geq i$.
\end{itemize}
\end{theorem}

\begin{proof}
(i) The inequalities in~(\ref{eq:cone}) are invariant under the multiplication of $\bm{\xi}$ by a positive number. If $p_j=0$, for $j>N$, they only involve $\xi_1,\dots,\xi_N$,  
and by~(\ref{eq:shadow}) they are equivalent to the same strict inequalities for the barycentric coordinates of the shadow, which by Lemma~\ref{lem:regions}(a) characterize the points of $A_i$, except those at which two of the ratios are minimal, a set of zero volume. 

(ii) For $u>0$, the number of indices $j$ such that $\xi_j/p_j<u$ has mean $\sum_j\Pp(\xi_j<p_ju)=\sum_j(1-e^{-p_ju})\leq u\sum_jp_j=u$ 
(since $1-e^{-x}\leq x$), hence it is finite with probability one; and the probability that there is no such index is $\prod_je^{-p_ju}=e^{-u}$, which tends to zero as $u\to\infty$. Thus, almost surely, there is a $u$ such that a finite, non-zero number of ratios are smaller than $u$, so that the minimum of the ratios exists; and it is attained by a single index, because the ratios have continuous distributions 
(two independent random variables with a probability density coincide with probability zero). 

(iii) and (iv) Under $\mu_\infty$, the ratio $\xi_i/p_i$ has the probability density $p_i\,e^{-p_is}$ 
[since $\Pp(\xi_i/p_i>s)=\Pp(\xi_i>p_is)=e^{-p_is}$], and conditionally on $\xi_i/p_i=s$, the probability that all the other ratios are larger than $s$ is, by independence, $\prod_{j\neq i}\Pp(\xi_j>p_js)=\prod_{j\neq i}e^{-p_js}=e^{-(1-p_i)s}$. Therefore
\begin{equation}
\mu_\infty\big(C_i(\mathbf{p})\cap\{T>u\}\big)=\int_u^\infty p_i\,e^{-p_is}\,e^{-(1-p_i)s}\,ds=p_i\,e^{-u}
\end{equation}
which gives (iv), and (iii) for $u=0$. 

(v) The Born probabilities of $D_N$ are $p_j/\sum_{k\leq N}p_k$, $j\leq N$, and the barycentric coordinates of $\pi_{N,\infty}(\bm{\xi})$ are $\xi_j/\sum_{k\leq N}\xi_k$, $j\leq N$. Their ratios are therefore the numbers $\xi_j/p_j$, $j\leq N$, up to a common factor, and their minimum is attained at $j=i$, as soon as $N\geq i$.
\end{proof}

\noindent Point (iii) is the infinite-dimensional form of~(\ref{eq:volumes}): the ratio of two volumes, which becomes meaningless in the limit, is replaced by the measure of a cone, which by Theorem~\ref{thm:projective}(ii) is the limit of these ratios, $\mu_\infty[C_i(\mathbf{p})]=\lim_{N\to\infty}\mu_N[A^{(N)}_i]=p_i$. The ``zero over zero'' of~\cite{AS2017b} is thus the limit of a ratio of two quantities vanishing at the same rate, to which the infinite membrane gives a direct meaning, independently of the normalization.

Point (v) strengthens the argument of~\cite{AS2014} (Sec.~14) recalled in the Introduction: the $N$-outcome measurements on the compressed states $D_N$, performed with the shadows of a same measurement interaction, do not merely have converging probabilities, but eventually all produce the same outcome. We also obtain an answer to the question of~\cite{AS2017b}: the law $\mu_\infty$ of the infinite membrane is invariant under the permutations of the outcomes, and the state breaks this symmetry by selecting the cones~(\ref{eq:cone}), whose weights are its components $p_i$ along the vertices: a weighted symmetry breaking over the hidden-measurement interactions, for an arbitrary number of outcomes.

\subsection{Uniqueness}
\label{subsec:unique}

In the EBR, the uniformity of the membrane is a hypothesis, from which the Born rule follows~\cite{AS2014}. Non-uniform membranes describe measurements which are neither classical nor quantum, and the uniform one emerges as an average over all the possible membranes, in what has been called a \emph{universal measurement}~\cite{AS2014,SdB2026b}. The following theorem shows that it is also the \emph{only} membrane which yields the Born rule for all the states.

\begin{theorem}[Uniqueness of the membrane]
\label{thm:unique}
\begin{itemize}
\item[]
\item[(a)] Let $\nu$ be a probability measure on $\triangle_{N-1}$, $N\geq 2$, which gives zero probability to the boundary of the simplex, i.e., such that $\nu(\lambda_j>0,\text{ for all }j)=1$. Suppose that $\nu$ reproduces the Born rule for all the states whose on-simplex particle lies in the interior of the simplex: for every $\mathbf{p}=(p_1,\dots,p_N)$, with $p_i>0$ and $\sum_ip_i=1$,
\begin{equation}
\nu(A_i)=p_i,\quad i=1,\dots,N
\label{eq:nuBornN}
\end{equation}
where the $A_i$ are the regions of Lemma~\ref{lem:regions} associated with the on-simplex particle whose barycentric coordinates are $\mathbf{p}$. Then $\nu=\mu_N$.
\item[(b)] Let $\nu$ be a probability measure on $\Rinf$, with $\nu(\xi_j>0,\text{ for all }j)=1$. Suppose that $\nu$ reproduces the Born rule for the states having a finite number of non-zero Born probabilities: for every $N$ and every $\mathbf{p}=(p_1,\dots,p_N,0,0,\dots)$, with $p_i>0$ and $\sum_{i\leq N}p_i=1$,
\begin{equation}
\nu\big[C_i(\mathbf{p})\big]=p_i,\quad i=1,\dots,N
\label{eq:nuBorn}
\end{equation}
Then all the shadows of $\nu$ are uniform, and consequently $\nu(C)=\mu_\infty(C)$ for every measurable cone $C$. If, moreover, $\lim_{n\to\infty}\frac{1}{n}\sum_{j\leq n}\xi_j = 1$, for $\nu$-almost every $\bm{\xi}$, then $\nu=\mu_\infty$.
\end{itemize}
\end{theorem}

\begin{proof}
(a) Since the regions $A_i$ fill the simplex and, by~(\ref{eq:nuBornN}), their probabilities sum to one, their intersections have zero probability. Hence, by Lemma~\ref{lem:regions}(a), up to a set of zero probability, $\bm{\lambda}\in A_N$ if and only if $\varrho_j>a_j$, for all $j<N$, where $\varrho_j=\lambda_j/\lambda_N\in(0,\infty)$ and $a_j=p_j/p_N$. Since $p_N=(1+\sum_{j<N}a_j)^{-1}$ and $p_j=a_j\,p_N$, the $a_j$ can be arbitrary positive numbers, every point of the interior of the simplex being the on-simplex particle of some state, and~(\ref{eq:nuBornN}) with $i=N$ gives
\begin{equation}
\nu\big(\varrho_1>a_1,\dots,\varrho_{N-1}>a_{N-1}\big)=\frac{1}{1+a_1+\dots+a_{N-1}},\quad\text{for all }a_1,\dots,a_{N-1}>0
\label{eq:ratios}
\end{equation}
and by~(\ref{eq:volumes}) the same holds for $\mu_N$. The sets $\{\varrho_j>a_j,\text{ for all }j<N\}$ form a family which is closed under intersection [the intersection of the sets associated with $(a_j)$ and $(b_j)$ is the set associated with $(\max\{a_j,b_j\})$], and which generates all the events concerning $(\varrho_1,\dots,\varrho_{N-1})$. Two probability measures which agree on such a family agree everywhere~\cite{Billingsley1995}. Hence, the ratios have the same joint distribution under $\nu$ and under $\mu_N$, and so has $\bm{\lambda}$, which is a function of them: $\lambda_N=(1+\sum_{k<N}\varrho_k)^{-1}$ and $\lambda_j=\varrho_j\lambda_N$, for $j<N$. 

(b) Fix $N$, and let $\nu_N$ be the distribution of the shadow $\pi_{N,\infty}(\bm{\xi})$ under $\nu$, which gives zero probability to the boundary of $\triangle_{N-1}$. By Theorem~\ref{thm:born}(i), whose characterization of the cones does not involve the probability law of $\bm{\xi}$, (\ref{eq:nuBorn}) says that $\nu_N$ gives the probability $p_i$ to the set of the points of $A_i$ at which $\lambda_i/p_i$ is the only minimal ratio, for $i=1,\dots,N$. These sets being disjoint, with probabilities summing to one, the points at which two ratios are minimal have zero probability, and $\nu_N$ satisfies~(\ref{eq:nuBornN}). By (a), $\nu_N=\mu_N$, and the last two statements follow from Theorem~\ref{thm:projective}(iv).
\end{proof}

\noindent It is remarkable that, for $N\geq 3$, the probability of a \emph{single} outcome, as a function of the state, suffices to reconstruct the membrane. Since the outcome probabilities depend linearly on the distribution of the measurement interactions, point (a) also implies that an average of membranes performed independently of the state reproduces the Born rule if and only if the average membrane is uniform, which complements the results on universal measurements. The infinite membrane is thus characterized in two independent ways, as the only distribution whose shadows are the uniform membranes (Theorem~\ref{thm:projective}), and as the only one compatible with the Born rule [Theorem~\ref{thm:unique}(b)], so that any other way of taking the limit of the EBR must lead to the same object.

\subsection{Which directions are states}

The cones~(\ref{eq:cone}) can be defined for any sequence $\mathbf{q}$ of non-negative numbers, and only depend on its direction. In finite dimension, every direction of the positive orthant is that of a state, $\sum_jq_jP_{a_j}/\sum_kq_k$. In infinite dimension, if $\sum_jq_j=\infty$, $\mathbf{q}$ is not the direction of any state, in the same way as a pseudo-eigenvector of position cannot be normalized: the relevant distinction is not between normalized and non-normalized vectors, but between \emph{normalizable} and \emph{non-normalizable directions}.\footnote{These are the directions of the cone of the positive operators commuting with $A$, issued from the zero operator, and not those of the generalized Bloch ball, issued from its center.} The measurement interactions are themselves non-normalizable directions, since $\sum_j\xi_j=\infty$.

\begin{proposition}[Normalizable and non-normalizable directions]
\label{prop:normalizable}
Let $\mathbf{q}=(q_1,q_2,\dots)$ be a sequence of non-negative numbers, not all equal to zero, and let $C_i(\mathbf{q})$ be the cones defined by~(\ref{eq:cone}), with $\mathbf{p}$ replaced by $\mathbf{q}$.
\begin{itemize}
\item[(a)] If $\sum_jq_j<\infty$, then $\mu_\infty$-almost every $\bm{\xi}$ belongs to exactly one of the cones $C_i(\mathbf{q})$, and $\mu_\infty[C_i(\mathbf{q})]=q_i/\sum_jq_j$, for every $i$.
\item[(b)] If $\sum_jq_j=\infty$, then $\mu_\infty[C_i(\mathbf{q})]=0$, for every $i$: for $\mu_\infty$-almost every $\bm{\xi}$, the infimum of the ratios $\xi_j/q_j$ is zero, and it is not attained, so that $\bm{\xi}$ belongs to none of the cones.
\end{itemize}
\end{proposition}

\begin{proof}
(a) The cones do not change if $\mathbf{q}$ is multiplied by a positive number, and we can apply Theorem~\ref{thm:born} to $\mathbf{p}=\mathbf{q}/\sum_jq_j$. (b) For $u>0$, the events $\{\xi_j<u\,q_j\}$ are independent, with probabilities $1-e^{-uq_j}$, whose sum is infinite when $\sum_jq_j=\infty$ 
(if infinitely many $u\,q_j$ are larger than one, infinitely many terms are larger than $1-e^{-1}$. Otherwise, all but finitely many terms are larger than $u\,q_j/2$, since $1-e^{-x}\geq x/2$ for $0\leq x\leq 1$). By the second Borel-Cantelli lemma~\cite{Billingsley1995}, according to which, if the probabilities of a sequence of independent events have an infinite sum, infinitely many of these events occur with probability one, infinitely many of the events $\{\xi_j<u\,q_j\}$ occur. This being true for every $u=1,\frac{1}{2},\frac{1}{3},\dots$, the infimum of the ratios is zero, and it cannot be attained since $\xi_j>0$ for all $j$.
\end{proof}

\noindent An example of case (b) is the center of the renormalized ball, $\mathbf{q}=(1,1,1,\dots)$, i.e., what would be the maximally mixed state: the outcome would be the facet closest to the measurement interaction, which does not exist, by Proposition~\ref{prop:where}(d). The infinite membrane thus produces an outcome exactly for the directions of the positive trace-class operators: the hidden-measurement mechanism distinguishes by itself the states from the ``non-normalizable states''.

\begin{figure}[t]
\centering
\includegraphics[width=0.74\textwidth]{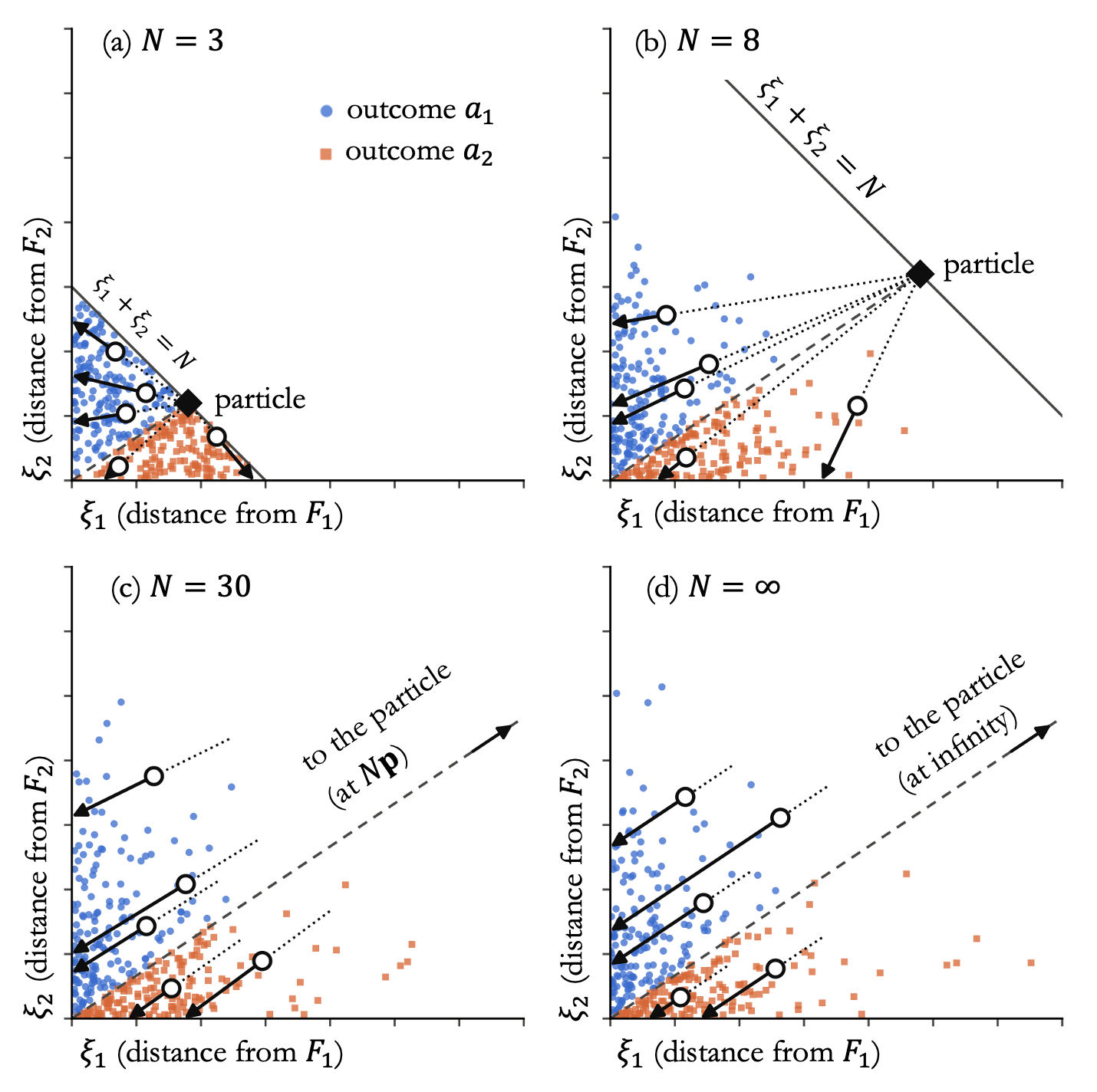} 
\caption{The uniform membrane in the renormalized ball, here for a state with Born probabilities $p_1=0.6$ and $p_2=0.4$, and an increasing number $N$ of outcomes. Distances $(\xi_1,\xi_2)$ from the first two facets (the vertical and the horizontal axis) of the shadows $N\pi_{N,\infty}(\bm{\xi})$ of the same measurement interactions $\bm{\xi}$ of the infinite membrane are shown. The dashed line separates the regions $A_1$ (round blue markers) and $A_2$ (square orange markers). (a) For $N=3$, the membrane is a triangle, with the particle on one of its edges. (b)-(c) As $N$ increases, the edge $\xi_1+\xi_2=N$ and the particle, at $(Np_1,Np_2)$, recede, whereas each disintegration point converges to its limit position; see Theorem~\ref{thm:projective}(iii). (d) In the limit, the distances become independent exponential variables, the regions are cones, and the particle is at infinity. For some measurement interactions (white markers), the lines of collapse are also shown. They pass through the particle in finite dimension, and become parallel in the infinite limit; see Proposition~\ref{prop:parallel}.}
\label{fig:renorm}
\end{figure}

\subsection{The particle recedes to infinity}
\label{subsec:recede}

By~(\ref{eq:xi}), the particle is at distance $\eta_j=Np_j$ from the facet $F_j$, which tends to infinity for every outcome of non-zero probability: in the renormalized ball, the particle is sent to infinity, in the direction $\mathbf{p}$, whereas in the unit ball the measurement interactions are sent to the center, and by Lemma~\ref{lem:regions}(c) the two can never be at a finite distance from each other. A point at infinity is nothing but a direction, and it is interesting to see what becomes of the geometry of the collapse. In finite dimension, by Lemma~\ref{lem:regions}(b), the outcome is determined by following the half-line issued from the particle and passing through the disintegration point, until it reaches a facet: all these half-lines converge at the particle; see Figs.~\ref{fig:distances}(b) and~\ref{fig:renorm}(a).

\begin{proposition}[The lines of collapse become parallel]
\label{prop:parallel}
Let $\mathbf{p}$ be the Born probabilities of the state $D$, and let $\bm{\xi}$ be a measurement interaction of the infinite membrane which satisfies Proposition~\ref{prop:where}(a) and (b) and belongs to one of the cones $C_i(\mathbf{p})$, as is the case for $\mu_\infty$-almost every $\bm{\xi}$ [Theorem~\ref{thm:born}(ii)]. For every sufficiently large $N$, consider, in the renormalized ball of radius $N-1$, the shadow of $\bm{\xi}$, at distances $\xi^{(N)}_j=N[\pi_{N,\infty}(\bm{\xi})]_j$ from the facets, the particle associated with the compressed state $D_N$, at distances $\eta^{(N)}_j=Np_j/\sum_{k\leq N}p_k$ from the facets, and the half-line issued from the particle and passing through the shadow, whose points are at distances
\begin{equation}
\xi^{(N)}_j(\sigma)=\xi^{(N)}_j+\frac{\sigma}{N}\left[\xi^{(N)}_j-\eta^{(N)}_j\right],\quad\sigma\geq 0
\label{eq:halfline}
\end{equation}
from the facets. Then, as $N\to\infty$:
\begin{itemize}
\item[(a)] $\xi^{(N)}_j(\sigma)\to\xi_j-\sigma\,p_j$, for every $j$ and every $\sigma\geq 0$;
\item[(b)] the value $\sigma_N$ of the parameter at which the half-line leaves the simplex converges to $T(\bm{\xi})=\min_j\xi_j/p_j$, and so does $N$ times the relative depth of the shadow, $N\,t[\pi_{N,\infty}(\bm{\xi})]\to T(\bm{\xi})$.
\end{itemize}
\end{proposition}

\begin{proof}
The distance from a facet being an affine function of the point, the point $\mathbf{x}+s(\mathbf{x}-\mathbf{y})$ of the half-line issued from $\mathbf{y}$ and passing through $\mathbf{x}$ is at distance $d(\mathbf{x},F_j)+s[d(\mathbf{x},F_j)-d(\mathbf{y},F_j)]$ from $F_j$, which gives~(\ref{eq:halfline}) with $s=\sigma/N$. (a) By Theorem~\ref{thm:projective}(iii), $\xi^{(N)}_j\to\xi_j$, so that $\xi^{(N)}_j/N\to 0$, whereas $\eta^{(N)}_j/N\to p_j$. (b)
The $j$-th distance~(\ref{eq:halfline}) vanishes for $\sigma/N=\omega_j/(1-\omega_j)$, where $\omega_j=\xi^{(N)}_j/\eta^{(N)}_j$, if $\omega_j<1$ (otherwise it never vanishes). This being an increasing function of $\omega_j$, the half-line leaves the simplex for $\sigma_N=N\,t_N/(1-t_N)$, where $t_N=\min_{j\leq N}\omega_j$ is the relative depth of the shadow. Now, $N\,t_N=\big(\frac{1}{N}\sum_{k\leq N}\xi_k\big)^{-1}\big(\sum_{k\leq N}p_k\big)\min_{j\leq N}\xi_j/p_j$, where the first two factors tend to one, and the third is equal to $T(\bm{\xi})$ for $N$ large enough. Hence $N\,t_N\to T(\bm{\xi})$, $t_N\to 0$, and $\sigma_N\to T(\bm{\xi})$.
\end{proof}

\noindent In the limit, the collapse of the membrane is therefore described by the half-line
\begin{equation}
\bm{\xi}(\sigma)=\bm{\xi}-\sigma\,\mathbf{p},\quad 0\leq\sigma\leq T(\bm{\xi})
\label{eq:parallel}
\end{equation}
The disintegration, which starts at the measurement interaction $\bm{\xi}$, propagates along the fixed direction $-\mathbf{p}$, i.e., away from the particle at infinity, until it reaches a facet, which is the facet $F_i$ such that $\bm{\xi}\in C_i(\mathbf{p})$, since the $i$-th coordinate of~(\ref{eq:parallel}) is the first to vanish. The half-lines associated with the different measurement interactions are now all parallel [Fig.~\ref{fig:renorm}(d)]: the particle acts like a lamp which, when it is moved farther and farther away, ends up illuminating the scene with parallel rays. The quantity $T(\bm{\xi})$ of Theorem~\ref{thm:born}(iv) is the length of the path of the disintegration, in units of $\mathbf{p}$, and its exponential law is the limit of that of the relative depth: by Lemma~\ref{lem:regions}(c), $\mu_N(N\,t>u)=(1-\frac{u}{N})^{N-1}\to e^{-u}$.

The deterministic fall of the particle onto the simplex, i.e., the map $D\mapsto\sum_iP_{a_i}DP_{a_i}$~\cite{AS2014}, survives the limit unchanged, at the scale of the states, whereas the collapse takes place at the scale of the measurement interactions, where the particle is only a direction. For a degenerate observable, the outcome associated with a spectral projection $P_I=\sum_{i\in I}P_{a_i}$ is the union of the cones $C_i(\mathbf{p})$, $i\in I$, of probability $\Tr DP_I$.  
The post-measurement state, however, is not determined by the cones: as in finite dimension, the transition to the L\"uders state $P_IDP_I/\Tr DP_I$ is described by an additional deterministic purification stage, which is part of the EBR modeling of degenerate measurements~\cite{AS2014,AS2017b}, and which differs from a non-degenerate measurement whose outcome is subsequently forgotten, which would produce the state $\sum_{i\in I}P_{a_i}DP_{a_i}/\Tr DP_I$.

\section{Measurements with a continuous spectrum}
\label{sec:continuous}

An observable with a continuous spectrum, like the position of a particle, has no eigenvectors, hence no vertices and no facets to which the above construction could be directly applied. However, as emphasized in~\cite{AS2017b}, every measurement with a finite number of outcomes, obtained by dividing the spectrum into a finite number of cells, is described by the EBR, each outcome being associated with an infinite-dimensional spectral projection. We can therefore follow what happens to these finite-dimensional membranes when the cells become smaller and smaller.

\subsection{Particle in a box}

Consider a particle in a box, $\mathcal{H}=L^2([0,L])$, with position observable $(A\psi)(x)=x\,\psi(x)$, and a state $D$ with Born density $p(x)$, i.e., ${\cal P}_D(B)=\Tr DP(B)=\int_Bp(x)\,dx$, where $P(B)$ is the multiplication by the characteristic function of the Borel set $B$. Dividing the box into $N$ cells $I_j=[(j-1)\epsilon,j\epsilon)$ of length $\epsilon=L/N$, the measurement with the $N$ outcomes ``the particle is in $I_j$'', of probabilities $p_j={\cal P}_D(I_j)$, is described by a uniform membrane with $N$ vertices and, in the renormalized ball, the on-simplex particle is at distances
\begin{equation}
\eta_j=Np_j=\frac{L}{\epsilon}\int_{I_j}p(y)\,dy\, \underset{N\to\infty}{\longrightarrow}\, L\,p(x)
\label{eq:eta-limit}
\end{equation}
from the facets, for almost every $x$, $j$ being the index of the cell containing $x$, since the mean value of $p$ over $I_j$ tends to $p(x)$. Contrary to Section~\ref{sec:purepoint}, the particle thus remains at a finite distance from the facets, its limit being described by $L\,p(x)$, the Born density relative to the uniform distribution, which would be that of the maximally mixed state, at the center of the ball. The renormalized ball now contains both the particle and the membrane, whose distances $\xi_j$ from the facets become a ``white noise'' of independent exponential variables over the cells; see Fig.~\ref{fig:rain}(a).

This is not yet sufficient to determine the outcome. By~(\ref{eq:rule}), the outcome minimizes $\xi_j/\eta_j$, and by Lemma~\ref{lem:regions}(c) this minimum is of order $1/N$. Since the $\eta_j$ are of order one, the outcome is decided among the few facets from which the disintegration point is at a distance of order $1/N=\epsilon/L$. Only the layer of the membrane which is closest to the facets matters, and to see it a second magnification is necessary. For this, we will consider the distances $\tau_j=\xi_j/\epsilon$ from the facets, in a ball of radius $(N-1)/\epsilon$.

\subsection{A Poisson rain}

The natural object to describe the limit of the bottom layer is a Poisson point process~\cite{Kingman1993}. Given a space $\mathcal{X}$ with a $\sigma$-finite measure $m$ without atoms (the \emph{intensity}), a Poisson point process on $\mathcal{X}$ with intensity $m$ is a random countable set $\Pi\subset\mathcal{X}$ such that the number of points of $\Pi$ in a measurable set $B$ is a Poisson random variable of mean $m(B)$, the numbers of points in disjoint sets being independent: a ``rain'' of points falling independently on $\mathcal{X}$, with density $m$. We will use the \emph{restriction theorem} (the points falling in disjoint sets form independent Poisson processes, with the restricted intensities) and the \emph{mapping theorem} (if $g:\mathcal{X}\to\mathcal{Y}$ is measurable and the image of $m$ by $g$ is $\sigma$-finite without atoms, $g(\Pi)$ is a Poisson process with this image intensity)~\cite{Kingman1993}.

\begin{proposition}[The bottom layer]
\label{prop:rare}
Let $(\xi_1,\dots,\xi_N)$ be the distances from the facets of a uniformly distributed point of the renormalized simplex, let $x_j$ be the center of the cell $I_j$, and $\tau_j=\xi_j/\epsilon$, with $\epsilon=L/N$. For $0\leq a<b\leq L$ and $c>0$, the number of points $(x_j,\tau_j)$ in the rectangle $[a,b)\times(0,c)$ converges in distribution, as $N\to\infty$, to a Poisson random variable of mean $(b-a)\,c$, and the numbers of points in disjoint rectangles become independent: the set of the points $(x_j,\tau_j)$ converges to a Poisson point process on $[0,L]\times(0,\infty)$ with intensity $dx\,d\tau$.
\end{proposition}

\begin{proof}
By Lemma~\ref{lem:uniform}, $\tau_j=(E_j/\epsilon)/(S_N/N)$, with $S_N/N\to 1$ almost surely. For the points $(x_j,E_j/\epsilon)$, the number of points in $[a,b)\times(0,c)$ is binomial, with about $(b-a)/\epsilon$ trials [the cells with center in $[a,b)$] of success probability $1-e^{-c\epsilon}\approx c\,\epsilon$, hence it converges to a Poisson variable of mean $(b-a)\,c$ (\emph{law of rare events}), and the same argument, applied to the multinomial distribution, shows that the counts in finitely many disjoint rectangles converge jointly to independent Poisson variables. The factor $S_N/N$ does not affect these limits: for $0<\delta<1$, with probability tending to one, $1-\delta<S_N/N<1+\delta$, and then the counts of the points $(x_j,\tau_j)$ and $(x_j,E_j/\epsilon)$ in disjoint rectangles $[a_k,b_k)\times[c_k,d_k)$ can only differ because of points $(x_j,E_j/\epsilon)$ lying in the strips $[a_k,b_k)\times[c(1-\delta),c(1+\delta))$, with $c=c_k$ or $c=d_k$, whose mean number tends to $\sum_k2\delta\,(b_k-a_k)(c_k+d_k)$, which is arbitrarily small for $\delta$ small enough. This joint convergence of the counts in finitely many rectangles (whose boundaries almost surely contain no point of the limit process) is the convergence in distribution of point processes~\cite{Kallenberg2017}.
\end{proof}

\begin{figure}[tp]
\centering
\includegraphics[width=0.8\textwidth]{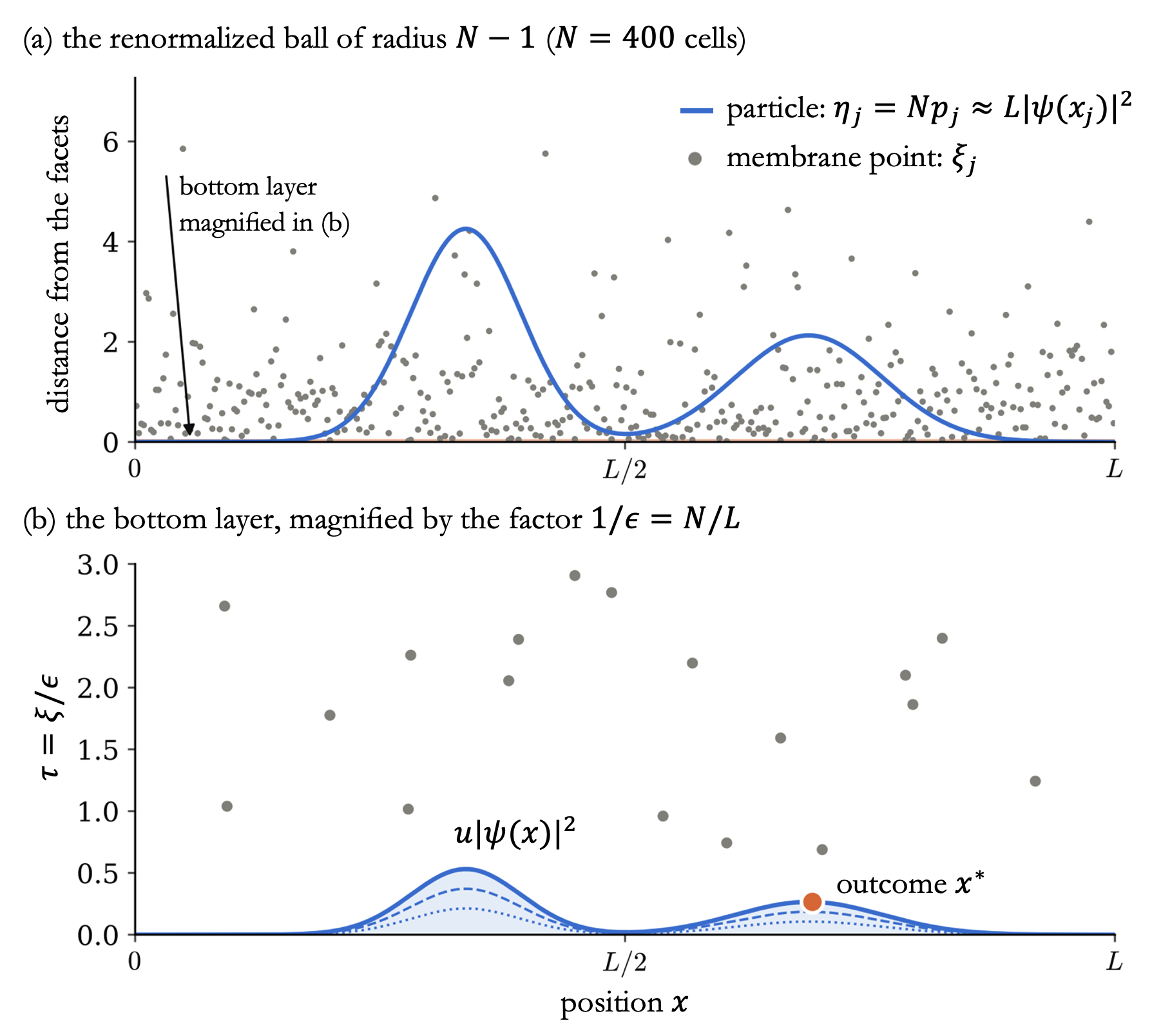} 
\caption{The membrane of a position measurement in a box of length $L$, here divided into $N=400$ cells of length $\epsilon=L/N$. (a) In the renormalized ball of radius $N-1$, the distances $\xi_j$ of the disintegration point of the membrane from the $N$ facets (grey dots) are approximately independent exponential variables, whereas those of the particle are $\eta_j\approx L\,p(x_j)$ (blue curve), where $p(x)=|\psi(x)|^2$ is the Born density. The outcome is the cell which minimizes $\xi_j/\eta_j$, and it only depends on the dots of the bottom layer (the thin strip along the horizontal axis, indicated by the arrow). (b) The bottom layer, magnified by the factor $1/\epsilon$: the dots now form a uniform Poisson rain in the plane $(x,\tau)$. The outcome $x^*$ is the abscissa of the point which minimizes $\tau/p(x)$, i.e., of the first point touched by the profile $u|\psi(x)|^2$ when $u$ increases from zero (dotted and dashed lines: the profile for earlier values of $u$).}
\label{fig:rain}
\end{figure}

\noindent The bottom layer of the membrane, suitably magnified, thus converges to a uniform rain of points on the product of the spectrum by a half-line; see Fig.~\ref{fig:rain}(b). As for the pure point spectrum, this convergence in distribution can be turned into a much more precise statement: the measurement interactions of all the finite-dimensional measurements can be read on a single realization of the rain.

\begin{theorem}[The membrane of a continuous spectrum]
\label{thm:rain}
Let $\Pi$ be a Poisson point process on $[0,L]\times(0,\infty)$ with intensity $dx\,d\tau$.
\begin{itemize}
\item[(i)] Let $I_1,\dots,I_N$ be a partition of $[0,L]$ into intervals, of lengths $\epsilon_1,\dots,\epsilon_N$, and let $\tau_j$ be the height of the lowest point of $\Pi$ above $I_j$, i.e., the smallest $\tau$ such that $(x,\tau)\in\Pi$, for some $x\in I_j$. Then the numbers $E_j=\epsilon_j\tau_j$ are independent $\Exp(1)$ random variables, so that
\begin{equation}
\xi_j=\frac{N\,\epsilon_j\,\tau_j}{\epsilon_1\tau_1+\dots+\epsilon_N\tau_N},\quad j=1,\dots,N
\end{equation}
are the distances from the facets of a uniformly distributed point of the renormalized simplex: they are a realization of the measurement interaction of the measurement with the $N$ outcomes $I_j$.
\item[(ii)] With this measurement interaction, the outcome of the measurement on an entity in the state $D$ is the cell $I_{j^*}$ which minimizes $\tau_j/\bar{p}_j$, where $\bar{p}_j={\cal P}_D(I_j)/\epsilon_j$ is the mean value of the Born density in the cell.
\item[(iii)] With probability one, there is a unique point $(x^*,\tau^*)$ of $\Pi$ which minimizes $\tau/p(x)$ (with $\tau/0=+\infty$). The probability that $x^*$ belongs to a set $B$ is $\int_Bp(x)\,dx$ (the Born rule), and $T=\tau^*/p(x^*)$ has the distribution $\Exp(1)$ and is independent of $x^*$.
\item[(iv)] If the Born density $p(x)$ is continuous on $[0,L]$, then, with probability one, $x^*\in I_{j^*}$ for every partition whose cells are small enough.
\end{itemize}
\end{theorem}

\begin{proof}
(i) By the restriction theorem, the points of $\Pi$ above the different cells form independent Poisson processes, so that the $\tau_j$ are independent, and the probability that $\tau_j>s$ is the probability that there is no point in $I_j\times(0,s)$, i.e., $e^{-\epsilon_js}$. Hence, $\epsilon_j\tau_j\sim\Exp(1)$, and the conclusion follows from Lemma~\ref{lem:uniform} and~(\ref{eq:xi}). 

(ii) By~(\ref{eq:rule}), with $\eta_j=N{\cal P}_D(I_j)$, the outcome minimizes $\xi_j/\eta_j$, which is proportional to $\epsilon_j\tau_j/{\cal P}_D(I_j)=\tau_j/\bar{p}_j$. 

(iii) Discarding the points with $p(x)=0$, the mapping $(x,\tau)\mapsto(x,s)$, with $s=\tau/p(x)$, transforms $\Pi$ into a Poisson process with intensity ${\cal P}_D(dx)\,ds$, by the mapping theorem, since the intensity of the set of the points mapped into $B\times(0,u)$ is $\int_Bdx\int_0^\infty d\tau\,\mathbf{1}\{\tau<u\,p(x)\}=u\,{\cal P}_D(B)$. For every $u>0$, the number of its points with $s<u$ is a Poisson variable of mean $u$: it is finite, and it is zero with probability $e^{-u}$, which tends to zero as $u\to\infty$. Hence, with probability one, there is a $u$ for which this number is finite and non-zero, so that there is a point of minimal $s$. This point is unique with probability one, since, by the mapping theorem, the values of $s$ form a Poisson process of unit intensity on the half-line, whose points are all distinct. Moreover, by the restriction theorem, the lowest transformed points above $B$ and above its complement $B^c$ have independent heights, with distributions $\Exp[{\cal P}_D(B)]$ and $\Exp[{\cal P}_D(B^c)]$, and the computation in the proof of Theorem~\ref{thm:born} gives that the probability that $x^*\in B$ and $T>u$ is ${\cal P}_D(B)\,e^{-u}$.

(iv) Let $\bar{p}$ be the step function equal, on each cell, to the mean value of the Born density in that cell [point (ii)]: $\bar{p}(x)=\bar{p}_j={\cal P}_D(I_j)/\epsilon_j$, for $x\in I_j$. Since $\bar{p}$ is constant on each cell, the outcome $I_{j^*}$ of (ii) is the cell containing the point of $\Pi$ which minimizes $\tau/\bar{p}(x)$, and we have to show that this point is $(x^*,\tau^*)$ when the cells are small. Let $M=(T+1)\,p_{\max}$, where $p_{\max}$ is the maximum of $p$, which also bounds $\bar{p}$. The points of $\Pi$ with $\tau\geq M$ have both $\tau/p(x)$ and $\tau/\bar{p}(x)$ larger than or equal to $T+1$, whereas the points with $\tau<M$ are finite in number, with distinct values of $\tau/p(x)$, the smallest one being $T$, at $(x^*,\tau^*)$. Since $p$ is uniformly continuous, $\bar{p}\to p$ uniformly when the size of the largest cell tends to zero, so that, at each of these finitely many points, $\tau/\bar{p}(x)\to\tau/p(x)$ [also when $p(x)=0$, both being, or tending to, $+\infty$]. Hence, when the cells are small enough, $(x^*,\tau^*)$ minimizes $\tau/\bar{p}(x)$ over the whole of $\Pi$.
\end{proof}

\noindent For a continuous spectrum, the infinite membrane is thus the Poisson rain $\Pi$,  
which contains the uniform membranes of all the measurements obtained by dividing the spectrum into cells, through the lowest points of the rain above the cells [point (i)] (a state-independent construction, which however differs from the shadow maps of Section~\ref{sec:infinite-membrane}), determines by itself an outcome $x^*$ distributed according to the Born density [point (iii)], and agrees with the outcomes of these membranes for fine enough partitions [point (iv), the analogue of Theorem~\ref{thm:born}(v)]. Geometrically [Fig.~\ref{fig:rain}(b)], the profile $u\,p(x)$ rises through the rain as $u$ increases, and the outcome is the abscissa of the first point it touches, the value $T$ of $u$ being the analogue of the length of the path~(\ref{eq:parallel}). An actual apparatus only registers a cell $I_j$, the state being then brought to $P(I_j)DP(I_j)/{\cal P}_D(I_j)$  
by the deterministic stage of the EBR for degenerate measurements~\cite{AS2014,AS2017b}, whereas $x^*$, to which no normalizable eigenvector is associated, belongs to the hidden level of the description, as the point of the spectrum at which the disintegration of the membrane reaches the boundary.

\subsection{General observables}

Our construction is not specific to position. Let $A$ be an arbitrary self-adjoint operator, with spectral measure $P(\cdot)$ on its spectrum $\sigma(A)$, ${\cal P}_D(B)=\Tr DP(B)$, and let $m$ be a $\sigma$-finite \emph{reference measure} on $\sigma(A)$ with respect to which every ${\cal P}_D$ has a density, ${\cal P}_D(B)=\int_Bp(x)\,m(dx)$, for instance $m={\cal P}_{D_0}$, with $D_0$ a state whose eigenvalues are all strictly positive (${\cal P}_{D_0}(B)=0$ implies $P(B)=0$, and the Radon-Nikodym theorem~\cite{Billingsley1995} applies). For the position, $m$ is the Lebesgue measure; for a pure point spectrum, $m$ can be the counting measure of the eigenvalues, $p$ being then the sequence of the Born probabilities. If $m$ is not finite on bounded sets (as for the eigenvalues $1/n$), the rain below must be understood as a \emph{Poisson random measure}~\cite{Kingman1993}; the transformed process of the proof has however the intensity ${\cal P}_D(dx)\,ds$, which is finite on $\sigma(A)\times(0,u)$, so that its lowest point is always well defined.

\begin{proposition}[The infinite membrane of a general observable]
\label{prop:general}
Let $\Pi$ be a Poisson point process on $\sigma(A)\times(0,\infty)$ with intensity $m(dx)\,d\tau$, and let $(x^*,\tau^*)$ be the point of $\Pi$ which minimizes $\tau/p(x)$
 (with the convention $\tau/0=+\infty$, as in Theorem~\ref{thm:rain}(iii)), and $T=\tau^*/p(x^*)$. Then:
\begin{itemize}
\item[(i)] $x^*$ and $T$ are well defined with probability one, the probability that $x^*\in B$ is ${\cal P}_D(B)$ for every Borel set $B$, and $T\sim\Exp(1)$ is independent of $x^*$;
\item[(ii)] the joint distribution of $(x^*,T)$ does not depend on the choice of the reference measure $m$;
\item[(iii)] if $m$ is the counting measure of a pure point spectrum, the heights $\xi_j$ of the lowest points of $\Pi$ above the eigenvalues $a_j$ are independent $\Exp(1)$ random variables, and $x^*=a_i$ if and only if $\bm{\xi}\in C_i(\mathbf{p})$: one recovers the infinite membrane of Definition~\ref{def:membrane};
\item[(iv)] for every partition of $\sigma(A)$ into Borel sets $B_1,\dots,B_N$ of non-zero probability, the numbers $E_k=\min\{\tau/p(x):(x,\tau)\in\Pi,\ x\in B_k\}\,{\cal P}_D(B_k)$ are independent $\Exp(1)$ random variables, and $x^*\in B_k$ if and only if $E_k/{\cal P}_D(B_k)$ is the smallest of the ratios: the measurement with the $N$ outcomes $B_k$ is described by a uniform membrane with $N$ vertices, as in~\cite{AS2017b}.
\end{itemize}
\end{proposition}

\begin{proof}
(i) and (iv) are proved as Theorem~\ref{thm:rain}(iii): the mapping $(x,\tau)\mapsto(x,\tau/p(x))$ transforms $\Pi$ into a Poisson process with intensity ${\cal P}_D(dx)\,ds$, the lowest points of which above the disjoint sets $B_k$ have independent heights with distributions $\Exp[{\cal P}_D(B_k)]$. 

(ii) The transformed process has intensity ${\cal P}_D(dx)\,ds$ whatever the reference measure: if $m$ is replaced by another measure $m'$, the density $p(x)$ changes, but not the measure $p(x)\,m(dx)={\cal P}_D(dx)$. 

(iii) For the counting measure, the points of $\Pi$ above a given eigenvalue form a Poisson process of unit intensity on the half-line, whose lowest point has the distribution $\Exp(1)$, independently for the different eigenvalues, and the other points can never be selected.
\end{proof}

\noindent  
Point (ii) shows that the construction is intrinsic as far as the outcome and the length $T$ are concerned. The rain itself depends on the choice of $m$, but different choices give equivalent descriptions: if $m'=g\,m$, with $g>0$, the map $(x,\tau)\mapsto(x,\tau/g(x))$ transforms a rain of intensity $m(dx)\,d\tau$ into one of intensity $m'(dx)\,d\tau$, while the density becomes $p/g$, so that $\tau/p(x)$, hence the outcome, is unchanged for every realization. 

By (iii), the pure point and continuous cases are particular cases of a single description, and by (iv) every finite measurement, described in~\cite{AS2017b} by a uniform membrane, can be obtained from this single object. Note however that the $E_k$ of (iv) depend on the state, through $p$: these membranes reproduce the outcome $x^*$ for every realization, but they are not state-independent, contrary to those of Theorem~\ref{thm:rain}(i), whose outcomes agree with $x^*$ only for fine enough partitions [Theorem~\ref{thm:rain}(iv), established for a continuous Born density].

\section{Discussion and conclusions}
\label{sec:discussion}

In this concluding section, we trace the separation between the states and the measurement interactions back to two normalizations of the operators, 
describe a dynamical reading of the uniform membrane as a race between exponential clocks, and briefly revisit the finite-outcome viewpoint of~\cite{AS2017b}.

\subsection{Inside or outside the Bloch ball?}

For every finite $N$, the REBR is the standard EBR in a different unit of length, so that the infinite membrane is entirely a product of its geometry. However, by Proposition~\ref{prop:nogo}, it does not live in $\mathcal{B}_\infty$. The reason lies in two normalizations of the operators: a state is a positive operator of unit \emph{trace}, a measurement interaction $\tilde{D}(\bm{\xi})$ of the infinite membrane a positive operator of unit \emph{mean eigenvalue}. In finite dimension, they are equivalent, since $\Tr\tilde{D}=N$ is the same as $\Tr(\tilde{D}/N)=1$, and a single simplex can contain both the particle and the disintegration point. In infinite dimension, they are incompatible, since a sequence of non-negative numbers cannot have both a finite sum and a non-zero mean.  

For a pure point spectrum, what they still have in common is the cone of the positive operators commuting with the measured observable, of which both the measurement interaction and the decohered state $\sum_iP_{a_i}DP_{a_i}$, represented by the on-simplex particle, are rays, and since the outcome only depends on the rays, this is all that is needed. For a continuous spectrum, where no non-zero multiplication operator is trace-class, the role of this cone is played by the positive measures on the spectrum, to which the Born probability measure ${\cal P}_D$ and the reference measure $m$ of the rain belong (Section~\ref{sec:continuous}).

\subsection{The membrane as an exponential race}
\label{subsec:race}

The representation of Lemma~\ref{lem:uniform} gives the measurement process an additional dynamical reading. If $\lambda_j=E_j/S_N$, then, by Lemma~\ref{lem:regions}(a), the outcome is the index $i$ which minimizes $\lambda_i/p_i$, i.e., since $S_N$ is common to all the ratios, the index which minimizes $T_j=E_j/p_j$, $j=1,\dots,N$. Since $\Pp(T_j>t)=e^{-p_jt}$, the $T_j$ are the ringing times of $N$ independent clocks, started at the same moment, the $j$-th clock running $p_j$ times faster than a standard one, and the outcome is the first clock to ring. Equivalently, the $E_j$ are the lengths of $N$ fuses, lit at the same moment and burning at the speeds $p_j$: the state does not affect the measurement interaction, i.e., the lengths, but only the speeds, exactly as in the EBR the membrane is the same for all the states. By the computation in the proof of Theorem~\ref{thm:born}(iii)--(iv), the $i$-th clock wins with probability $p_i$, and the time $T^*=\min_jT_j$ of the first ring is an $\Exp(1)$ variable, independent of the outcome.

The fuses are directly related to the lines of collapse. The vector $\mathbf{X}(t)=\mathbf{E}-t\,\mathbf{p}$ of the unburnt lengths moves in the direction $-\mathbf{p}$, and the first fuse is consumed at $T^*=S_N\,t(\bm{\lambda})$. Since $\mathbf{E}=S_N\bm{\lambda}$ and $\sum_jX_j(t)=S_N-t>0$, for $t\leq T^*$, the normalization of $\mathbf{X}(t)$ is the point with barycentric coordinates
\begin{equation}
\frac{\mathbf{X}(t)}{\sum_jX_j(t)}=\bm{\lambda}+\frac{t}{S_N-t}\,(\bm{\lambda}-\mathbf{p})
\label{eq:crack}
\end{equation}
which, as $t$ increases from $0$ to $T^*$, moves along the half-line issued from the particle and passing through $\bm{\lambda}$, from $\bm{\lambda}$ to the point $\mathbf{f}$ where it leaves the simplex [for $t=T^*$, (\ref{eq:crack}) is $(\bm{\lambda}-t(\bm{\lambda})\,\mathbf{p})/(1-t(\bm{\lambda}))$, i.e., $\mathbf{f}$, by the proof of Lemma~\ref{lem:regions}(a)]: \emph{the line of collapse of the membrane is the shadow, on the simplex, of the burning fuses}. In the renormalized ball, $\xi_j=E_j/(S_N/N)$, and in the limit the shadow and the fuses coincide, along~(\ref{eq:parallel}). The race is thus neither an alternative to the membrane, nor an additional hypothesis: it is what the membrane becomes when the radius of the generalized Bloch ball tends to infinity, and since only the rays matter, it is already exact for finite $N$.

This dynamical reading raises a number of further questions, which go beyond the infinite-dimensional limit of the EBR and will be addressed in a separate article: whether independent and identically distributed clocks other than the exponential ones can generate the uniform membrane; how the race behaves under coarse-graining, i.e., for degenerate observables, and how it relates to the Lüders rule; what its connection is with the random-utility models of choice of mathematical psychology; and how the measurement interaction, if regarded as an additional hidden variable attached to the entity, should be updated after a measurement.

\subsection{Revisiting the finite-outcome viewpoint}

In~\cite{AS2017b}, it was suggested, albeit speculatively, that the measurement interactions might supervene only when the potentiality of the entity is constrained by a finite number of possible outcomes. The present results show that, as far as the mechanism is concerned, this is not required: the hidden-measurement process is well defined for infinitely many outcomes, discrete or continuous, through an explicit and probabilistically consistent limit of the finite-dimensional constructions, and it produces an outcome after the disintegration has propagated along the line of collapse~(\ref{eq:parallel}) over a length $T$ whose law, $\Exp(1)$, depends neither on the state nor on the number of outcomes [Theorems~\ref{thm:born}(iv) and~\ref{thm:rain}(iii)]. Whether an actual apparatus can realize a measurement with infinitely many possible outcomes, of vanishing individual probabilities, on the other hand, concerns the realization of the mechanism, not its logical consistency. 

Note that, as a probability space, the infinite membrane is isomorphic to an interval with the Lebesgue measure, as in the hidden-measurement models with infinitely many outcomes of Coecke~\cite{Coecke1995}, but it is here the canonical limit of the hidden-measurement spaces of the EBR, together with their geometry: facets, cones, shadows and lines of collapse. 

Finally, the measurement interactions become simpler in the infinite limit: their distances from the facets become independent, and for a continuous spectrum independent in disjoint regions of the spectrum, so that the fluctuations of the measurement context which are responsible for the quantum probabilities appear as a noise which is local in the spectrum of the measured observable.
\\
%\section*{Statements and Declarations}
%\noindent\textbf{Funding.} The author did not receive support from any organization for the submitted work.
%\noindent\textbf{Competing interests.} The author has no relevant financial or non-financial interests to disclose.
%\noindent\textbf{Data availability.} Data sharing is not applicable to this article, as no datasets were generated or analysed during the current study.

\noindent\textbf{Use of generative AI.} The article was developed by the author drawing also on a dialogue with the artificial intelligence assistant Claude (Anthropic). ChatGPT (OpenAI) was subsequently used for critical review of the exposition. The author has reviewed and verified all arguments, proofs and references, and takes full responsibility for the content of the publication.

\end{document}